\documentclass{aa}
\usepackage{textcomp}
\usepackage{graphicx}
\usepackage{amsmath}
\usepackage{amssymb}
\usepackage{bm}
\usepackage{upgreek}
\usepackage{IEEEtrantools}
\usepackage{multirow}
\usepackage[dvipsnames]{xcolor}
\usepackage[colorlinks=true,allcolors=blue,urlcolor=blue]{hyperref}
\usepackage{txfonts}
\usepackage[normalem]{ulem} 
\usepackage{booktabs}
\usepackage{longtable}
\usepackage{natbib}
\usepackage{subcaption}

\newcommand{\eqn}[1]{\text{Equation~\ref{#1}}}
\newcommand{\sect}[1]{\text{Section~\ref{#1}}}
\newcommand{\fig}[1]{\text{Figure~\ref{#1}}}
\newcommand{\tab}[1]{\text{Table~\ref{#1}}}

\newcommand{\balder}{\texttt{Balder}}
\newcommand{\multitd}{\texttt{Multi3D}}

\newcommand{\marcs}{\texttt{MARCS}}
\newcommand{\nefertiti}{\texttt{NEFERTITI}}

\newcommand{\teff}{T_{\mathrm{eff}}}
\newcommand{\lgg}{\log{g}}
\newcommand{\lgeps}[1]{A(\mathrm{#1})}
\newcommand{\xfe}[1]{\mathrm{[#1/Fe]}}
\newcommand{\feh}{\mathrm{[Fe/H]}}
\newcommand{\vmic}{\xi_{\text{mic;1D}}}
\newcommand{\nm}{\mathrm{nm}}
\newcommand{\dex}{\mathrm{dex}}
\newcommand{\kms}{\mathrm{km\,s^{-1}}}

\newcommand{\kelvin}{\mathrm{K}}

\begin{document} 

\title{Revisiting inelastic Cu+H collisions 
and the non-LTE Galactic evolution of copper} 

\author{S.~Caliskan\inst{\ref{uu1}},
A.~M.~Amarsi\inst{\ref{uu1}},
M.~Racca\inst{\ref{SU}},
I.~Koutsouridou\inst{\ref{Firenze}},
P.~S.~Barklem\inst{\ref{uu1}},
K.~Lind\inst{\ref{SU}},
S.~Salvadori\inst{\ref{Firenze}}
}

\authorrunning{Caliskan et al.}

\institute{\label{uu1}Theoretical Astrophysics, 
Department of Physics and Astronomy,
Uppsala University, Box 516, SE-751 20 Uppsala, Sweden\\
\email{sema.caliskan@physics.uu.se}
\and
\label{SU}Department of Astronomy, Stockholm University, AlbaNova University Centre, Stockholm, Sweden
\and
\label{Firenze}Dipartimento di Fisica e Astronomia, Universit\'{a} degli Studi di Firenze, Largo E. Fermi 1, 50125, Firenze, Italy
}

\abstract{The Galactic evolution of copper remains poorly understood, partly due to the strong departures from local thermodynamic equilibrium (LTE) affecting \ion{Cu}{I} lines. A key source of uncertainty in non-LTE modelling is the treatment of inelastic Cu+H collisions. We present new rate coefficients based on a combined asymptotic LCAO (linear combination of atomic orbitals) and free electron model approach, which show significant differences from previous calculations. Applying these updated rates to non-LTE stellar modelling, we find reduced line-to-line scatter and improved consistency between metal-poor dwarfs and giants. Our non-LTE analysis reveals a strong upturn in the [Cu/Fe] trend towards lower $\mathrm{[Fe/H]}<-1.7$.  We show that this may reflect the interplay between external enrichment of Cu-rich material of the Milky Way halo at low metallicities, and metallicity-dependent Cu yields from rapidly rotating massive stars. This highlights the unique diagnostic potential of accurate Cu abundances for understanding both stellar and Galactic evolution.}

\keywords{atomic processes --- radiative transfer --- 
line: formation --- Stars: abundances --- Galaxy: abundances}

\date{Received 24 February 2025 / Accepted 28 March 2025}
\maketitle

\section{Introduction}
\label{introduction}

Copper (Cu) is a relatively abundant\footnote{$\lgeps{Cu}\equiv\log{N_{\mathrm{Cu}}/N_{\mathrm{H}}}+12=4.18$ in the Sun \citep{Asplund_2021A&A...653A.141A}.}, odd ($Z=29$) iron peak element that is an important diagnostic tool for constraining supernova nucleosynthesis and chemical evolution models. It has been suggested that Cu can be produced through various nucleosynthesis processes in different astrophysical sites, including explosive nucleosynthesis in core-collapse and Type Ia supernovae, the weak s-process in massive stars, and the main s-process in AGB stars \citep{Bisterzo_2004MmSAI..75..741B}. However, the exact contributions of each astrophysical site and their roles in the galactic chemical evolution of Cu remain unclear and are subjects of ongoing debate. Moreover, Cu is a so-called ``killing element'' \citep{Salvadori_2019MNRAS.487.4261S}, meaning it is not produced by Pair Instability Supernovae (PISN), a type of supernova explosion that leaves no remnant and whose existence is still debated \citep[e.g.][]{Schulze_2024A&A...683A.223S}, although indirect hints \citep{Pagnini_2023MNRAS.521.5699P}. So, an under-abundance of Cu and other killing elements relative to iron in metal-poor stars could be a signature of massive first stars that exploded as PISN \citep{Aguado_2023MNRAS.520..866A,Koutsouridou_2024ApJ...962L..26K}.

Early studies, such as \cite{Sneden_1991A&A...246..354S}, observed that in the metal-poor regime, the copper-to-iron ratio ([Cu/Fe]) increases with metallicity ([Fe/H]\footnote{$\mathrm{[X/H]}\equiv\lgeps{X}_{\text{star}}-
\lgeps{X}_{\text{Sun}}$.}). This was later confirmed by \cite{Bisterzo_2004MmSAI..75..741B,Bisterzo_2005NuPhA.758..284B}, and \cite{Bihain_2004A&A...423..777B} who found linear [Cu/Fe] trends at subsolar metallicities, consistent with a metallicity-dependent weak s-process nucleosynthesis.

However, these [Cu/Fe] trends are biased by departures from local thermodynamic equilibrium (LTE). Early works by \cite{Bihain_2004A&A...423..777B} and \cite{Bonifacio_2010A&A...524A..96B} already indicated that accurate Cu abundances require non-LTE corrections, particularly for UV lines. This was also suggested by \cite{Roederer_2014ApJ...791...32R} and \cite{Roederer_2018ApJ...857....2R}, who demonstrated that using \ion{Cu}{I} lines in LTE underestimates the Cu abundance, namely the [Cu/Fe] ratios determined from \ion{Cu}{II} lines are higher than those determined from \ion{Cu}{I} lines, due to overionisation of the neutral minority species.

The first non-LTE calculations for \ion{Cu}{I} were by \cite{Shi_2014ApJ...782...80S} and \cite{Yan_2015ApJ...802...36Y, Yan_2016A&A...585A.102Y}, who showed that for metal-poor dwarfs the non-LTE corrections yield larger [Cu/Fe] ratios and a much flatter [Cu/Fe] trend than in LTE. Later, these works were expanded by \cite{Shi_2018ApJ...862...71S}, who confirmed that there is still a linear increase of [Cu/Fe] at very low metallicities, but the trend is much more flattened. \cite{Andrievsky_2018MNRAS.473.3377A} calculated and applied non-LTE abundance corrections to a sample of both metal-poor dwarfs and giants and found even larger non-LTE corrections (up to +1 dex) such that the [Cu/Fe] vs [Fe/H] trend almost disappears in non-LTE.

But non-LTE studies are only as reliable as the atomic data that go into them, encapsulated in the so-called ``model atom''. Indeed, non-LTE modelling of stellar atmospheres requires detailed information on all significant radiative and collisional processes. In particular, the inelastic hydrogen collisions (excitation or de-excitation, as well as ion-pair production or mutual neutralization) are found to be a major source of uncertainty for non-LTE studies of late-type stars \citep[e.g.][]{Barklem_2011A&A...530A..94B, Barklem_2016A&ARv..24....9B}.

All of the afore-mentioned non-LTE modelling studies for Cu used the so-called Drawin formula for inelastic hydrogen collisions \citep{Steenbock_1984A&A...130..319S,Lambert_1993PhST...47..186L}. This often overestimates collisional rates by several orders of magnitude, due to its classical approximation that does not account for the necessary physics behind inelastic collisions with hydrogen, which is of quantum nature \citep{Barklem_2011A&A...530A..94B, Barklem_2016A&ARv..24....9B}. Detailed full quantum calculations provide the most accurate rate coefficients for hydrogen collisions \citep{Barklem_2016A&ARv..24....9B}, but are limited to a few simple systems such as Li, Na, and Mg \citep{Belyaev_2003A&A...409L...1B, Belyaev_1999PhRvA..60.2151B,Belyaev_2010PhRvA..81c2706B,Guitou_2011JPhB...44c5202G,Belyaev_2012PhRvA..85c2704B}, due to being extremely computationally demanding and time-consuming. Therefore, simplified methods like the ``asymptotic models'' described by \cite{Belyaev_2013A&A...560A..60B} and \cite{Barklem_2016PhRvA..93d2705B} are increasingly used to provide inelastic hydrogen collision data for a wide range of elements of astrophysical interest. These models offer a simplified yet physically motivated description of the quantum mechanical processes and collision dynamics. Comparisons with available full-quantum data show good agreement for the largest rate coefficients \citep{Barklem_2016A&ARv..24....9B}. Additionally, laboratory measurements of mutual-neutralisation processes of Li,
Na, and Mg showed that the asymptotic model successfully reproduces experimental branching coefficients \citep{Barklem_2021ApJ...908..245B, Grumer_2022PhRvL.128c3401G, Grumer_2023PhRvL.130b9901G}.

Asymptotic model data for inelastic Cu + H collisions have only become available very recently \citep{Belyaev_2021MNRAS.501.4968B}. \cite{Xu_2022ApJ...936....4X} incorporated these data into the model atom from \cite{Shi_2014ApJ...782...80S}, and found that these updates resulted in less severe non-LTE corrections and consequently lower Cu abundances, making the dependence of [Cu/Fe] on metallicity stronger.   

In this paper, we present a new set of Cu + H rate coefficients (\sect{collisions}), calculated based on the asymptotic model approach described in \cite{Barklem_2016PhRvA..93d2705B}. We combine these data with rates calculated using the free-electron model by \cite{Kaulakys_1991JPhB...24L.127K}, as this may lead to a more comprehensive description of hydrogen collisions \citep{Amarsi_2018A&A...616A..89A,2024PhRvA.109e2820S}. We explore the impact of these new Cu + H collision data on non-LTE modelling (\sect{methodnon-LTE}) and then apply non-LTE abundance corrections to literature data to determine the [Cu/Fe] versus [Fe/H] trend (\sect{gce}).  We use this to astrophysically validate the new collisional data, and then discuss what the non-LTE abundances tell us about the possible cosmic origins of Cu, before summarizing the key points of the paper (\sect{conclusion}).

\section{Inelastic hydrogen collisions} \label{collisions}

\subsection{Asymptotic (LCAO) model} \label{methodLCAO}

\begin{table*}[]
\footnotesize
\center
\caption{\label{tab:input_Cu} Asymptotic states for Cu + H included in the calculations.}
\begin{tabular*}{\textwidth}{l @{\extracolsep{\fill}} rcccrrrclr}
\toprule
$\mathrm{State}_A$ & $2S_A+1$ & $ L_A$ & $  n_A$ & $  l_A$ & $ E_j^\mathrm{Cu}$ [eV]& $E_j$[eV] & $  E_{lim}$ [eV]& $N_{eq}$ & $  \mathrm{State}_c$  & $ G^{S_A L_A}_{S_c L_c}$ \\ \midrule
\multicolumn{11}{l}{\textbf{Covalent states:} Cu(($^{2S_c+1}$ $L_c$) $nl$ $^{2S_A+1}$ $L_A$)+H(1s $^{2}\mathrm{{S}}$) } \\ \midrule
$ 3d^{10}4s~^2\mathrm{S}$   & 2 &  0 & 4 & 0 & 0.000000 & 0.000000 & 7.726380 & 1 & Cu$^+$ $3d^{10}~^1\mathrm{S}$ & 1.0000 \\
$ 3d^{9}4s^2~^2\mathrm{D}$ & 2 & 2 & 4 & 0 & 1.490259 &  1.490259 & 10.534471 & 2 & Cu$^+$ $3d^{9}4s~^3\mathrm{D}$ & 0.866 \\
$ 3d^{9}4s^2~^2\mathrm{D}$ & 2 & 2 & 4 & 0 & 1.490259 &  1.490259 & 10.982770 & 2 & Cu$^+$ $3d^{9}4s~^1\mathrm{D}$  & 0.500 \\
$ 3d^{10}4p~^2\mathrm{P^o}$   & 2 & 1 & 4 & 1 & 3.806427 & 3.806427  & 7.726380 & 1 &  Cu$^+$ $3d^{10}~^1\mathrm{S}$& 1.0000 \\
$ 3d^{9}4s4p~^4\mathrm{P^o}$ & 4 & 1 & 4 & 1 & 4.922749 & 4.922749 & 10.534471 & 1 &  Cu$^+$ $3d^{9}4s~^3\mathrm{D}$  & 1.0000 \\
$ 3d^{9}4s4p~^4\mathrm{F^o}$   & 4 & 3 & 4 & 1 & 5.122793 & 5.122793  & 10.534471 & 1 &  Cu$^+$ $3d^{9}4s~^3\mathrm{D}$  & 1.0000 \\
$ 3d^{10}5s~^2\mathrm{S}$   & 2 & 0 & 5 & 0 & 5.348335 & 5.348335  & 7.726380 & 1 & Cu$^+$ $3d^{10}~^1\mathrm{S}$ &  1.0000 \\
$ 3d^{9}4s4p~^4\mathrm{D^o}$   & 4 & 2 & 4 & 1 & 5.471163 &5.471163 & 10.534471 & 1 &  Cu$^+$ $3d^{9}4s~^3\mathrm{D}$ & 1.0000 \\

$ 3d^{9}4s4p~(^3\mathrm{P})~^2\mathrm{F^o}$   & 2 & 3 & 4 & 1 & 5.509004 &5.509004 & 10.534471 & 1 &  Cu$^+$ $3d^{9}4s~^3\mathrm{D}$  & 0.500 \\
$ 3d^{9}4s4p~(^3\mathrm{P})~^2\mathrm{F^o}$   & 2 & 3 & 4 & 1 & 5.509004 &5.509004 & 10.982770 & 1 &  Cu$^+$ $3d^{9}4s~^1\mathrm{D}$  & 0.866 \\
$ 3d^{9}4s4p~(^3\mathrm{P})~^2\mathrm{P^o}$   & 2 & 1 & 4 & 1 & 5.685898 &5.685898 & 10.534471 & 1 &  Cu$^+$ $3d^{9}4s~^3\mathrm{D}$  & 0.500 \\
$ 3d^{9}4s4p~(^3\mathrm{P})~^2\mathrm{P^o}$   & 2 & 1 & 4 & 1 & 5.685898 &5.685898 & 10.982770 & 1 &  Cu$^+$ $3d^{9}4s~^1\mathrm{D}$  & 0.866 \\
$ 3d^{9}4s4p~(^3\mathrm{P})~^2\mathrm{D^o}$ & 2 & 2 & 4 & 1 & 5.777459 &5.777459 & 10.534471 & 1 &  Cu$^+$ $3d^{9}4s~^3\mathrm{D}$ & 0.500 \\
$ 3d^{9}4s4p~(^3\mathrm{P})~^2\mathrm{D^o}$ & 2 & 2 & 4 & 1 & 5.777459 &5.777459 & 10.982770 & 1 &  Cu$^+$ $3d^{9}4s~^1\mathrm{D}$ & 0.866 \\
$ 3d^{10}5p~^2\mathrm{P^o}$ &  2 & 1 & 5 & 1 & 6.122718 &6.122718 & 7.726380 & 1 &  Cu$^+$ $3d^{10}~^1\mathrm{S}$ & 1.0000 \\
$ 3d^{10}4d~^2\mathrm{D}$ & 2 & 2 & 4 & 2 & 6.191685 &6.191685 & 7.726380 & 1 &  Cu$^+$ $3d^{10}~^1\mathrm{S}$& 1.0000 \\
$ 3d^{10}6s~^2\mathrm{S}$ & 2 & 0 & 6 & 0 & 6.552410 &6.552410 & 7.726380 & 1 & Cu$^+$ $3d^{10}~^1\mathrm{S}$ & 1.0000 \\
$ 3d^{10}6p~^2\mathrm{P}$ & 2 & 1 & 6 & 1 & 6.802430 &6.802430 & 7.726380 & 1 & Cu$^+$ $3d^{10}~^1\mathrm{S}$ & 1.0000 \\
$ 3d^{10}5d~^2\mathrm{D}$ & 2 & 2 & 5 & 2 & 6.867466 &6.867466 & 7.726380 & 1 &  Cu$^+$ $3d^{10}~^1\mathrm{S}$ & 1.0000 \\
$ 3d^{10}4f~^2\mathrm{F^o}$  & 2 & 3 & 4 & 3 & 6.872150 &6.872150 & 7.726380 & 1 &  Cu$^+$ $3d^{10}~^1\mathrm{S}$ & 1.0000 \\
$ 3d^{10}7s~^2\mathrm{S}$   & 2 & 0 & 7 & 0 & 7.026355 &7.026355  & 7.726380 & 1 & Cu$^+$ $3d^{10}~^1\mathrm{S}$ & 1.0000 \\
$ 3d^{9}4s4p~(^1\mathrm{P})~^2\mathrm{F^o}$ & 2 & 3 & 4 & 1 & 7.057848 &7.057848 & 10.534471 & 1 &  Cu$^+$ $3d^{9}4s~^3\mathrm{D}$ & 0.866 \\
$ 3d^{9}4s4p~(^1\mathrm{P})~^2\mathrm{F^o}$ & 2 & 3 & 4 & 1 & 7.057848 &7.057848 & 10.982770 & 1 &  Cu$^+$ $3d^{9}4s~^1\mathrm{D}$ & -0.500 \\
$ 3d^{9}4s4p~(^1\mathrm{P})~^2\mathrm{P^o}$ & 2 & 1 & 4 & 1 & 7.069245 &7.069245 & 10.534471 & 1 &  Cu$^+$ $3d^{9}4s~^3\mathrm{D}$ & 0.866 \\
$ 3d^{9}4s4p~(^1\mathrm{P})~^2\mathrm{P^o}$ & 2 & 1 & 4 & 1 & 7.069245 &7.069245 & 10.982770 & 1 &  Cu$^+$ $3d^{9}4s~^1\mathrm{D}$ & -0.500 \\
$ 3d^{9}4s4p~(^1\mathrm{P})~^2\mathrm{D^o}$ & 2 & 2 & 4 & 1 & 7.125026 &7.125026 & 10.534471 & 1 &  Cu$^+$ $3d^{9}4s~^3\mathrm{D}$ & 0.866 \\
$ 3d^{9}4s4p~(^1\mathrm{P})~^2\mathrm{D^o}$ & 2 & 2 & 4 & 1 & 7.125026 &7.125026 & 10.982770 & 1 &  Cu$^+$ $3d^{9}4s~^1\mathrm{D}$ & -0.500 \\
$ 3d^{10}7p~^2\mathrm{P^o}$   & 2 & 1 & 7 & 1 & 7.162831 &7.162831  & 7.726380 & 1 &  Cu$^+$ $3d^{10}~^1\mathrm{S}$  & 1.0000 \\
$ 3d^{10}6d~^2\mathrm{D}$ & 2 & 2 & 6 & 2 & 7.177968 & 7.177968 & 7.726380 & 1 &  Cu$^+$ $3d^{10}~^1\mathrm{S}$ &  1.0000 \\
$ 3d^{10}5f~^2\mathrm{F^o}$ & 2 & 3 & 5 & 3 & 7.179832 & 7.179832  & 7.726380 & 1 &  Cu$^+$ $3d^{10}~^1\mathrm{S}$  & 1.0000 \\
$ 3d^{10}5g~^2\mathrm{G}$ & 2 & 4 & 5 & 4 & 7.181677 &7.181677  & 7.726380 & 1 &  Cu$^+$ $3d^{10}~^1\mathrm{S}$ &  1.0000 \\
$ 3d^{10}8s~^2\mathrm{S}$ & 2 & 0 & 8 & 0 & 7.261618 &7.261618  & 7.726380 & 1 & Cu$^+$ $3d^{10}~^1\mathrm{S}$ &  1.0000 \\
$ 3d^{10}7d~^2\mathrm{D}$ & 2 & 2 & 7 & 2 & 7.346072 &7.346072  & 7.726380 & 1 &  Cu$^+$ $3d^{10}~^1\mathrm{S}$ &  1.0000 \\
$ 3d^{10}6f~^2\mathrm{F^o}$ & 2 & 3 & 6 & 3 & 7.347252& 7.347252  & 7.726380 & 1 &  Cu$^+$ $3d^{10}~^1\mathrm{S}$  & 1.0000 \\ \midrule
\\
\multicolumn{5}{l}{\textbf{Ionic states:} {$\mathrm{Cu}^+ (^{2S_A+1} L_A) +\mathrm{H}^-$}(1s$^2$ $^{2}{S}$) } &
$ E_j^{\mathrm{Cu}^+}$ [eV]&  $ E_j$[eV] & \multicolumn{4}{c}{}\\ \midrule
 Cu$^+$ $3d^{10}~^1\mathrm{S}$ & 1 & 0   & - & - & 7.726380 &  6.972185 & \multicolumn{3}{c}{} \\
 Cu$^+$ $3d^{9}4s~^3\mathrm{D}$ & 3  & 2 & - & - & 10.534471 & 9.780276 & \multicolumn{3}{c}{} \\
 Cu$^+$ $3d^{9}4s~^1\mathrm{D}$ &  1  & 2 & - & - & 10.982770 & 10.228575 & \multicolumn{3}{c}{} \\
\bottomrule
\end{tabular*}
\end{table*}

\begin{figure}
\centering
\includegraphics[width=\hsize]{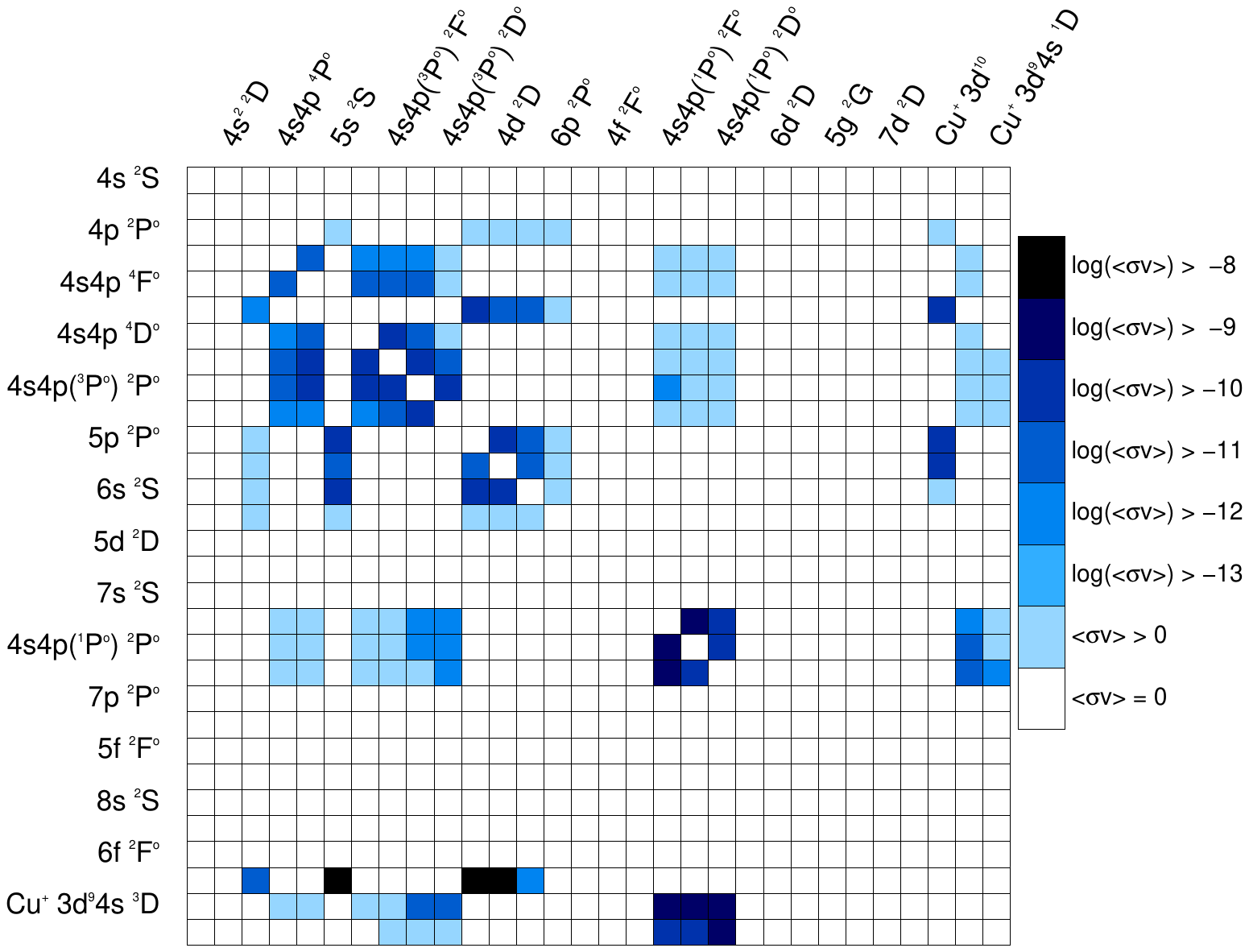}
\caption{Heat map of the rate coefficient matrix (in cm$^3$ s$^{-1}$) for inelastic Cu + H and Cu$^+$ + H$^-$ collisions at temperature T = 6000 K, calculated with the LCAO asymptotic model.
\label{fig:rates}}
\end{figure}

We performed calculations of the (de-)excitation and charge transfer processes of \ion{Cu}{I} through inelastic hydrogen collisions using the asymptotic method and codes described in \cite{Barklem_2016PhRvA..93d2705B, Barklem_2017PhRvA..95f9906B}. This method considers that during a collision, the valence electron from the atom of interest (\ion{Cu}{I} in this case) has a probability of tunneling to the H atom at a certain internuclear distance, known as an avoided ionic crossing. This process results in a Cu-H quasi-molecule with covalent and ionic molecular configurations. At the avoided crossing, the interaction between ionic and covalent configurations allows the electron from the Cu atom either to remain with the hydrogen atom, leading to ion-pair production (or mutual neutralization in the reverse process), or to tunnel back to a different covalent molecular state, resulting in excitation or deexcitation. The potentials and couplings resulting from the quasi-molecule and these covalent-ionic interactions are calculated using the asymptotic LCAO (linear combination of atomic orbitals) approach. The collisional cross-sections and rate coefficients for the various transitions are then computed with the multi-channel Landau-Zener model.

We present the covalent (Cu + H) and ionic (Cu$^+$ + H$^-$) asymptotic states considered in the calculations, along with the necessary input data, in \tab{tab:input_Cu}. We included all states up to 7.347 eV above the ground state of \ion{Cu}{I}, corresponding to 27 states. The energies for these levels are taken from the NIST Atomic Spectra Database (NIST ASD; \citealt{NIST_ASD}). To account for all ionic states appearing as core states among the included configurations, three ionic states were included as shown in \tab{tab:input_Cu}. Since Cu has $3d^9$ and $3d^{10}$ configurations, it is necessary to include both the $3d^{10}$ and $3d^{9}4s$ cores. Moreover, the states with $3d^9$ configurations can couple to both the triplet ($3d^{9}4s$ $^3\mathrm{D}$) and singlet ($3d^{9}4s$ $^1\mathrm{D}$) ionic states, so we included both.

Since these $3d^9$ states involve two different cores, we calculated the parentage coefficients ($|G^{S_A L_A}_{S_c L_c}|$) for each of them. This is important since the parentage coefficients are in the expression of the wavefunction for the Cu + H quasi-molecule based on the active electron on Cu and on the ground state of neutral hydrogen, which is used to calculate the potentials and couplings. A more detailed description of the calculations can be found in the \sect{appendix}. As a consequence of including all ionic states, the parentage coefficients satisfy the normalisation condition when summing over the different cores.

After testing several intervals, the internuclear distance for which the potential energies are calculated was set between R/a$_0$ = 3 and 200, since there were no avoided ionic crossings occurring at shorter and larger internuclear distances. The collisional dynamics calculations are performed for centre of mass energies between 10$^{-10}$ and 100 eV. We note that the asymptotic LCAO model only accounts for couplings between states corresponding to a one-electron transition. So two-electron transitions are excluded here. Additionally, the calculations are carried out in LS coupling, so fine-structure is not included. 

Rate coefficients for all the transitions between states listed in \tab{tab:input_Cu} are calculated for the temperatures from 1000 K to 20~000~K with steps of 1000 K. In \fig{fig:rates}, the rate coefficients at 6000 K are presented in a matrix form. This temperature is assumed for the following discussion. The lower diagonal of the matrix represents downward processes (de-excitation and mutual neutralisation), and the upper diagonal represents upward processes (excitation and ion pair production). 

The largest rate coefficients (on the order of $10^{-8}$ cm$^3$ s$^{-1}$) are for the mutual neutralisation processes involving the first ionic state  $3d^{10}~^1\mathrm{S}$, with the largest one to the final state Cu($3d^{10} 5s$ $^2\mathrm{S}$) + H(1s $^{2}\mathrm{S}$) with $4.5 \times 10^{-8}$ cm$^3$ s$^{-1}$. They are followed by mutual neutralisation between the other two ionic states $3d^9 4s$ $^3\mathrm{D}$ and $3d^9 4s$ $^1\mathrm{D}$ to the final states $3d^9 4s 4p$ $^2\mathrm{F}$, $^2\mathrm{P}$, and $^2\mathrm{D}$, with rates on the order of $10^{-9}$ cm$^3$ s$^{-1}$. It is found in previous works on asymptotic model calculations that the largest rates correspond to charge transfer processes, since this process involves only one transition (an electron moving from the Cu atom to the H atom), while (de-)excitation processes require two transitions (the electron moving back to the Cu atom in a different energy state) \citep{Barklem_2011A&A...530A..94B}. Finally, the other large rates involve (de-)excitation transitions between $3d^9 4s 4p~(^1\mathrm{P})$ $^2\mathrm{F}$, $^2\mathrm{P}$ and $^2\mathrm{D}$ states. This is consistent with previous studies \citep[e.g.][]{amarsi_2019A&A...625A..78A, Grumer_2020A&A...637A..28G}, that find that asymptotic LCAO calculations give largest rates for transitions involving states with intermediate and moderately-high energy values.

We also see that the rate coefficients for transitions involving the lowest two energy states $3d^{10}4s$ $^2\mathrm{S}$ and $3d^94s^2$  $^2\mathrm{D}$ are zero.  For these lowest energy states the corresponding avoided ionic crossings happen at very short internuclear distances, beyond the regime of the present model. Correspondingly, the transitions involving highly excited states (e.g. from $3d^{10} 7p$ $^2\mathrm{P}$ to $3d^{10} 6f$ $^2\mathrm{F}$) also give zero rate coefficients. These high-lying states have avoided crossings at large distances with very small couplings, which are thus traversed diabatically, and the electron transfer mechanism is not important.

Finally, the transitions between states with $3d^9$ and $3d^{10}$ configurations give zero rates in this model because these are two-electron processes. It is expected that two-electron transitions have a smaller probability of occurring since they require the simultaneous transition of two electrons, therefore giving too small rate coefficients to be of astrophysical relevance.

\subsection{Free electron model} \label{methodkaulakys}

\begin{figure*}
    \begin{center}
        \includegraphics[scale=0.56]{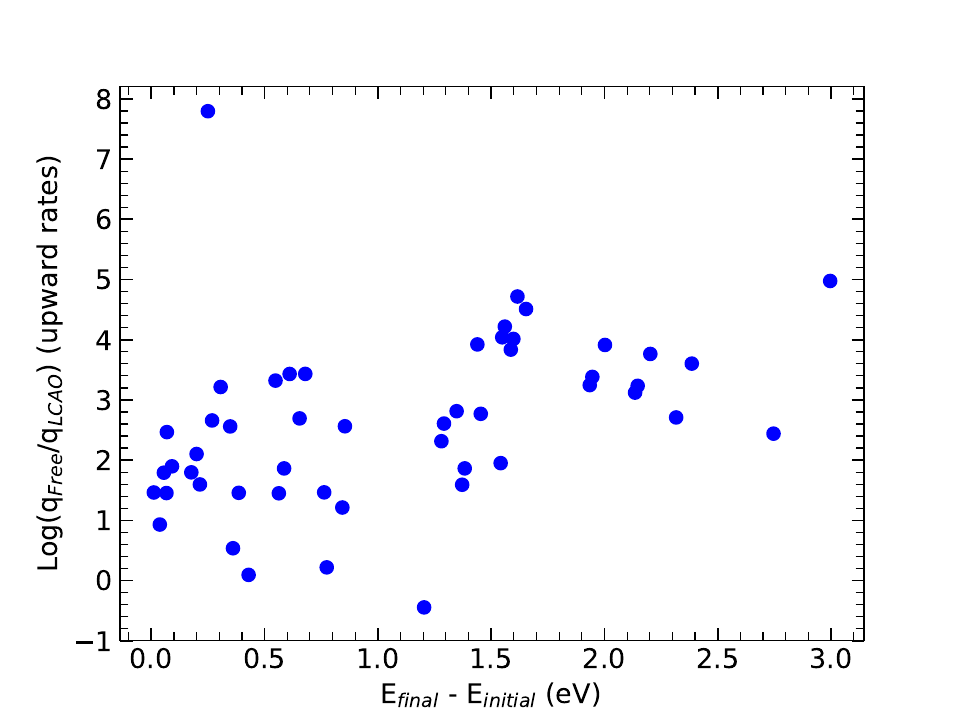}
        \includegraphics[scale=0.56]{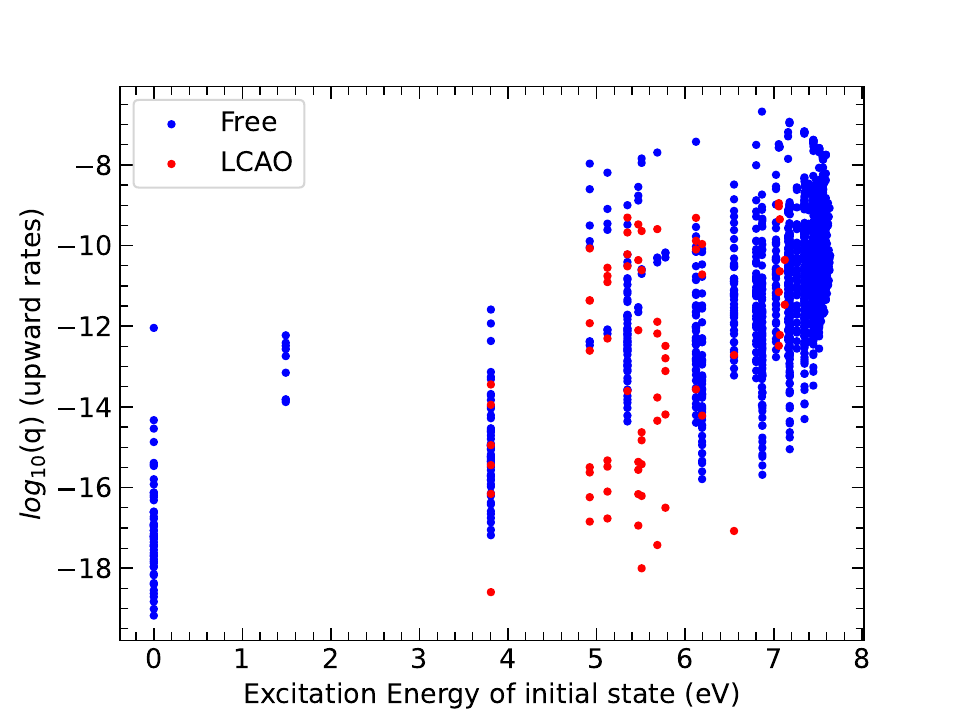}
        \caption{Left panel: Ratio of the non-zero rate coefficients (denoted as q) from the asymptotic LCAO and free electron model calculations, as a function of the transition energy, i.e. the difference between the energy of the final and initial state (for the states in \tab{tab:input_Cu}). 
        Right panel: Rate coefficients (q) from the extended asymptotic LCAO and free electron model calculations, as a function of the excitation energy of the initial state (for states up to 7.645 eV). For both panels, the rates are for the excitation and ion-pair production processes at 6000 K.}
        \label{fig:rates_Kaul_vs_ion}
    \end{center}
\end{figure*}

The asymptotic LCAO models predict very low or zero rate coefficients for transitions involving highly excited levels (near or above the relevant ionic limit) and for states where the avoided crossing occurs at long and short internuclear distances. These transitions are dominated by interaction mechanisms other than radial coupling at avoided ionic crossings, which are not accounted for in the asymptotic model \citep{Barklem_2011A&A...530A..94B}. The free electron model is suitable for these cases \citep{Barklem_2016A&ARv..24....9B}. Since this model does not include ionic configurations, it does not account for couplings at avoided ionic crossings. Therefore, as the two models describe different interaction mechanisms, by adding the coefficients from the asymptotic LCAO model and the free electron model, we achieve a more comprehensive description of the overall physics governing hydrogen collisions 
\citep{Amarsi_2018A&A...616A..89A,2024PhRvA.109e2820S}.

The rate coefficients were evaluated using the code from \cite{Barklem_2017ascl.soft01005B}, which is based on the free electron model by \cite{Kaulakys_1991JPhB...24L.127K}. In this model, momentum transfer between hydrogen (the perturber) and the Rydberg electron of the target atom, which is treated as a free electron, is considered. The code implements the analytical expression for the cross-section for a $nl \rightarrow n'l'$ transition (see Equations 6 to 8 in \citealt{Kaulakys_1991JPhB...24L.127K}) based on the impulse approximation. As for the asymptotic LCAO model, fine-structure is not included in the rate calculations. Furthermore, the original model from Kaulakys does not account for transitions between different spin states. The code follows the recipe from \cite{Osorio_2015A&A...579A..53O} to calculate the cross-sections redistributed between different spin states, which is based on the scattering length approximation (see Equations 8 to 12 in \citealt{Osorio_2015A&A...579A..53O}).
All the Rydberg states for \ion{Cu}{I} present in the model atom (see \sect{atom}) were included in the free electron model calculations.

\fig{fig:rates_Kaul_vs_ion} shows a comparison of the upward rate coefficients at 6000 K, obtained with the asymptotic LCAO and free electron models. We can see that for the transitions generating non-zero rates, the free electron model predicts systematically larger rate coefficients than the asymptotic LCAO model, for the states that were included in the LCAO calculation (in \tab{tab:input_Cu}). On average, the rate coefficients calculated with the free electron model are two to three orders of magnitude larger. Additionally, these differences are increasing with the transition energy. These results seem to be consistent with previous comparisons of the rate coefficients by the asymptotic LCAO and free electron models. \cite{Amarsi_2018A&A...616A..89A} also found larger rate coefficients with the free electron model for O + H collisions, and a similar trend with transition energy (see their Figure 5). It should be noted that the free electron model does not provide rate coefficients for the charge transfer processes. Since these processes were proven to be important by previous works as well as the present asymptotic LCAO model calculations, this confirms the importance of including the rate coefficients from the asymptotic LCAO model calculations.

For a better comparison of the rate coefficients from the free electron and asymptotic LCAO models, we performed an asymptotic LCAO calculation for an extended set of \ion{Cu}{I} states up to 7.645 eV, including Rydberg series up to $n$ = 15 and $l$ = 4. The resulting rate coefficients as well as the ones from the free electron model are shown in \fig{fig:rates_Kaul_vs_ion} (right panel). From this figure, we can see that the free electron model produces non-zero rate coefficients for transitions from the ground state and first excited state, while the asymptotic LCAO model predicts no transitions from these states. This is also the case for transitions from highly excited states (i.e. above 7.125 eV) for which the asymptotic LCAO model again predicts zero rates. This suggests that the mechanism described by the LCAO model, namely electron transfer at avoided ionic crossings, is not the dominant mechanism for these transitions which occur at too short or long range, and highlights the importance of adding the rate coefficients from the free electron model to avoid underestimating the hydrogen collisions in our atmospheric models. Moreover, the fact that the asymptotic LCAO model predicts zero rates for states above 7.125 eV motivates our choice of states included in the asymptotic model calculation (presented in \tab{tab:input_Cu}).

\subsection{Comparison with literature} \label{resultsBelyaev}

\begin{figure}
\centering
\includegraphics[width=\hsize]{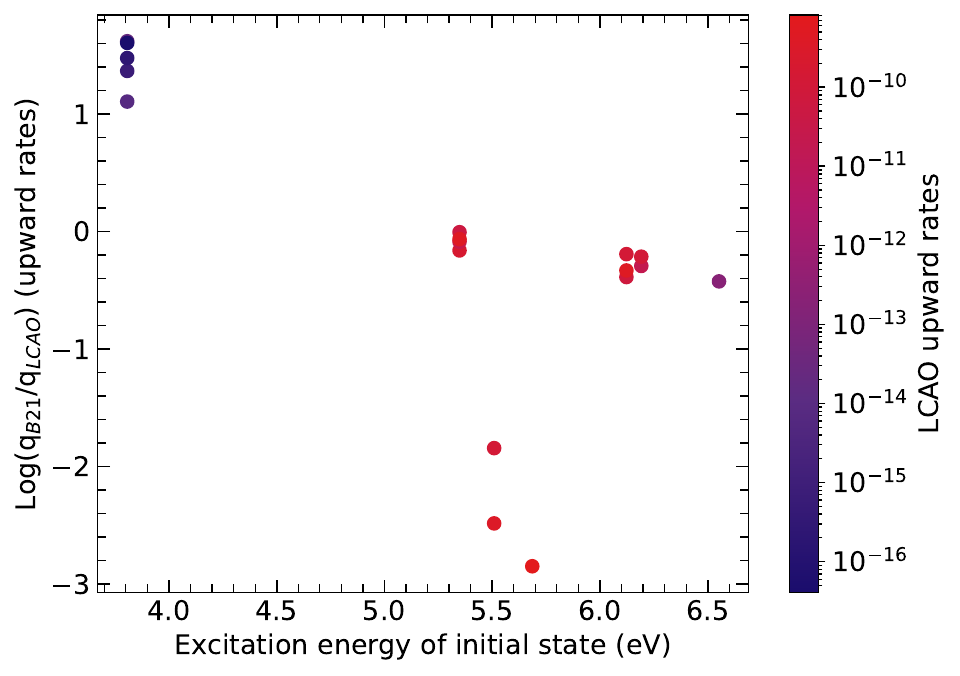}
\caption{Comparison of the collapsed rate coefficients from \cite{Belyaev_2021MNRAS.501.4968B} ({q$\mathrm{_{B21}}$}) and from our asymptotic LCAO model presented in \sect{methodLCAO} (q$\mathrm{_{LCAO}}$) for the excitation and ion-pair production processes at 6000 K as a function of the excitation energy of the initial state. These rates are for the 18 levels included in the B21 calculation.
\label{fig:rates_belyaev_vs_ion}}
\end{figure}

To assess the reliability of our newly calculated rate coefficients, we compare them with existing ones from the literature. For our rate coefficients from the free electron calculation, we could not find results based on the free electron approach for \ion{Cu}{I} in the literature. Concerning the asymptotic LCAO rates, the most recent Cu+H collision data based on an asymptotic model approach including similar physics come from \cite[][hereafter B21]{Belyaev_2021MNRAS.501.4968B}, who calculated rate coefficients for (de-)excitation and charge transfer processes for 17 low-lying doublet states of \ion{Cu}{I} and one ionic state, including fine-structure effects.

Comparing our asymptotic LCAO rate coefficients with those of B21 provides a detailed check on both our calculations and theirs, as well as serving as a verification of the asymptotic model approach for Cu+H collisions. There are, however, some differences in the methods used. While our calculations use an LCAO approach to compute the coupling coefficients, B21 employ a semi-empirical method. Additionally, our LCAO model is based on the LS coupling scheme, whereas B21 use a transformation from LS to JJ representation to incorporate fine-structure effects. Another distinction is that our calculations include three ionic states to account for couplings with the \ion{Cu}{I} states with $3d^9$ configurations and the $3d^9 4s$ $^3\mathrm{D}$ and $^1\mathrm{D}$ ionic states, while B21 include only the first ionic state, $3d^{10}$ $^1\mathrm{S}$. Additionally, the calculations of B21 include two-electron transition processes, while our asymptotic LCAO model only accounts for one-electron transitions. Finally, because B21 rates include fine-structure effects while the LCAO rates do not, in order to have a consistent comparison, we collapsed the upward B21 rates to LS rates following Equation 3 in \cite{Lind_2024ARA&A..62..475L} and applying it to rate coefficients instead. The reverse downward rates were then obtained through the detailed balance relations.

By comparing the rates for transitions included in both studies, we observe that some transitions produce zero rates in our calculations but yield non-zero values in B21 results. These are the transitions from $3d^{10} 4p$ $^2\mathrm{P}$ and $3d^{10} 5s$ $^2\mathrm{S}$ to the ground state $3d^{10} 4s$ $^2\mathrm{S}$, as well as all the transitions involving the $3d^9 4s^2$ $^2\mathrm{D}$ state. The reasons behind these zero rates from our asymptotic LCAO calculation are discussed in \sect{methodLCAO}. Additionally, our rates for all two-electron transitions are zero, in contrast to the non-zero values reported by B21, as expected given our different approaches to transition mechanisms.

\fig{fig:rates_belyaev_vs_ion} compares the non-zero rates (denoted as q) at 6000 K from both studies. They both identify the largest rate as corresponding to the same transition, namely the mutual neutralisation process $\text{Cu}^+ + \text{H}^- \rightarrow \text{Cu} (3d^{10} 5s\, ^2S) + \text{H}$, with agreement within $\sim$16$\%$. While some rates agree closely (with log(q$_{\mathrm{B21}}$/q$_{\mathrm{LCAO}}$) $\approx$ 0), others differ by 1 to 3 orders of magnitude. For instance, our LCAO rates for the important transitions between the $3d^9 4s 4p$ $^2\mathrm{L}$ states are 2 to 3 orders of magnitude larger than those from B21. In contrast, the B21 rates for transitions from the  $3d^{10} 4p$ $^2\mathrm{P}$ state to all states with $3d^{10}$ configurations are over 10 times larger. We discuss the key differences further in the context of non-LTE modelling in \sect{non-LTE} below.

\section{Non-LTE modelling} \label{methodnon-LTE}

\begin{figure*}
    \begin{center}
        \includegraphics[scale=0.32]{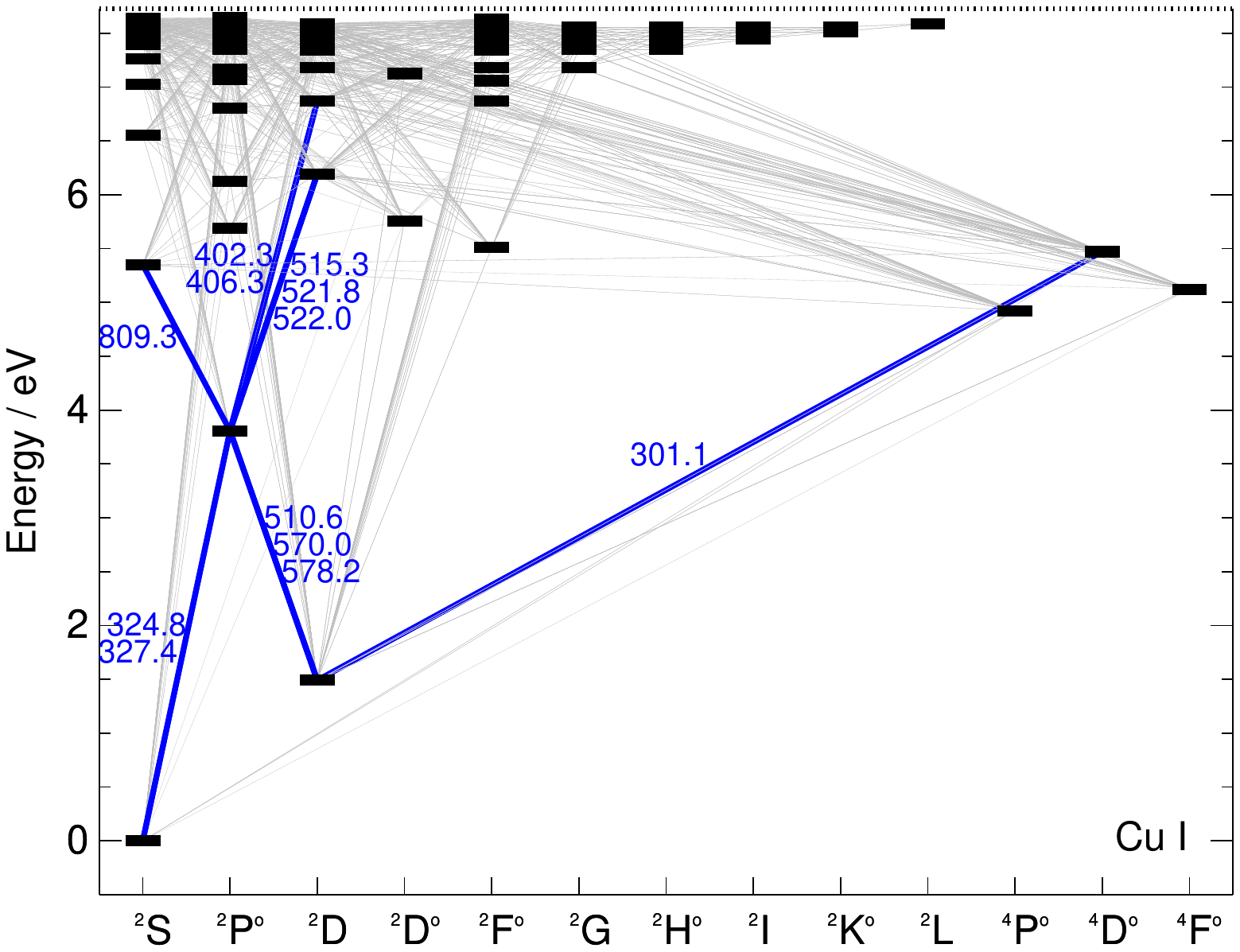}
        \includegraphics[scale=0.32]{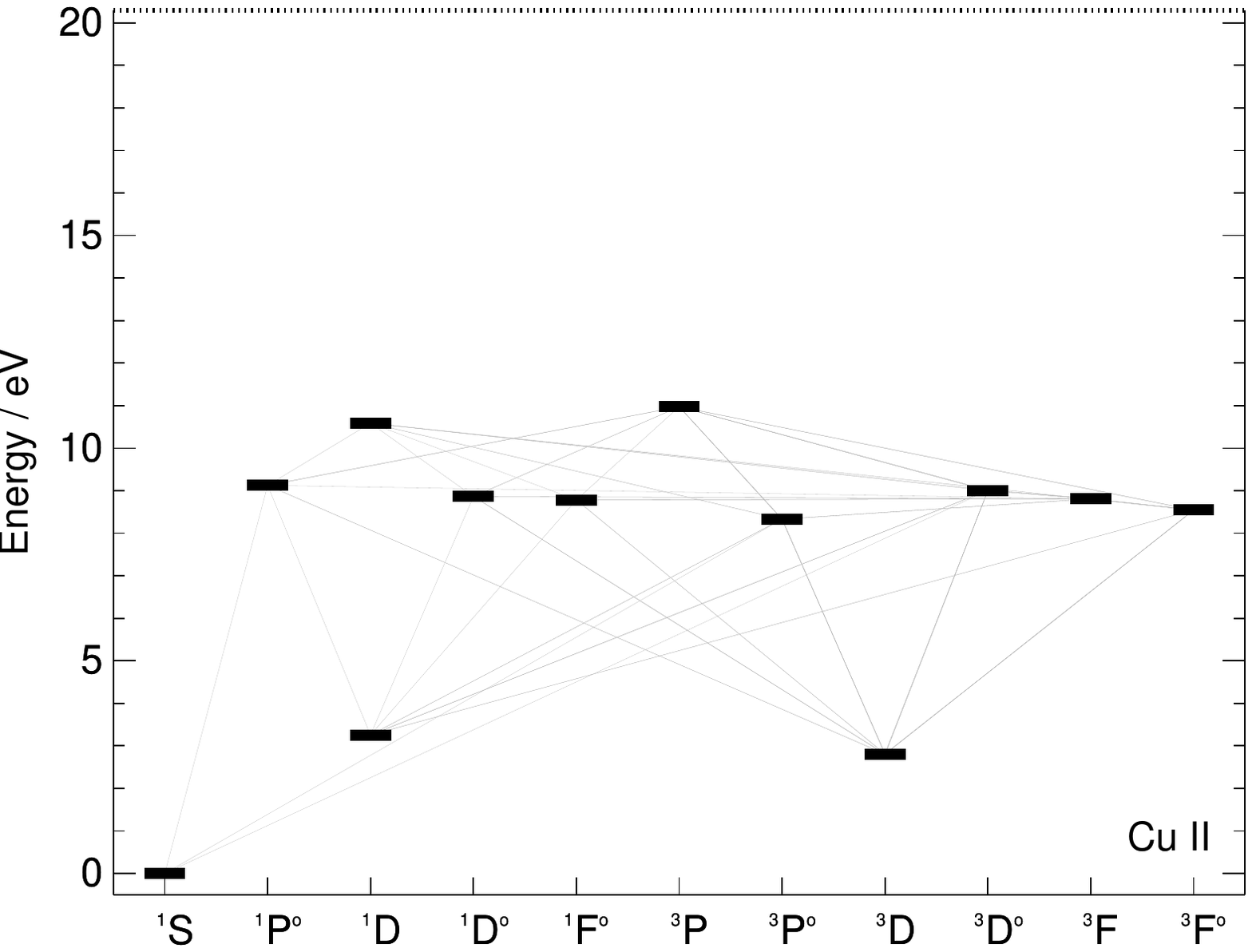}
        \caption{Grotrian diagrams of \ion{Cu}{I} (left panel) and \ion{Cu}{II} (right panel) representing the model atom described in \sect{atom}. The lines highlighted in blue are the 12 \ion{Cu}{I} lines analyzed in this paper (wavelengths in vacuum). The dotted horizontal line represents the ionization limit.}
        \label{fig:modelatom}
    \end{center}
\end{figure*}

\begin{figure*}
    \begin{center}
        \includegraphics[scale=0.355]{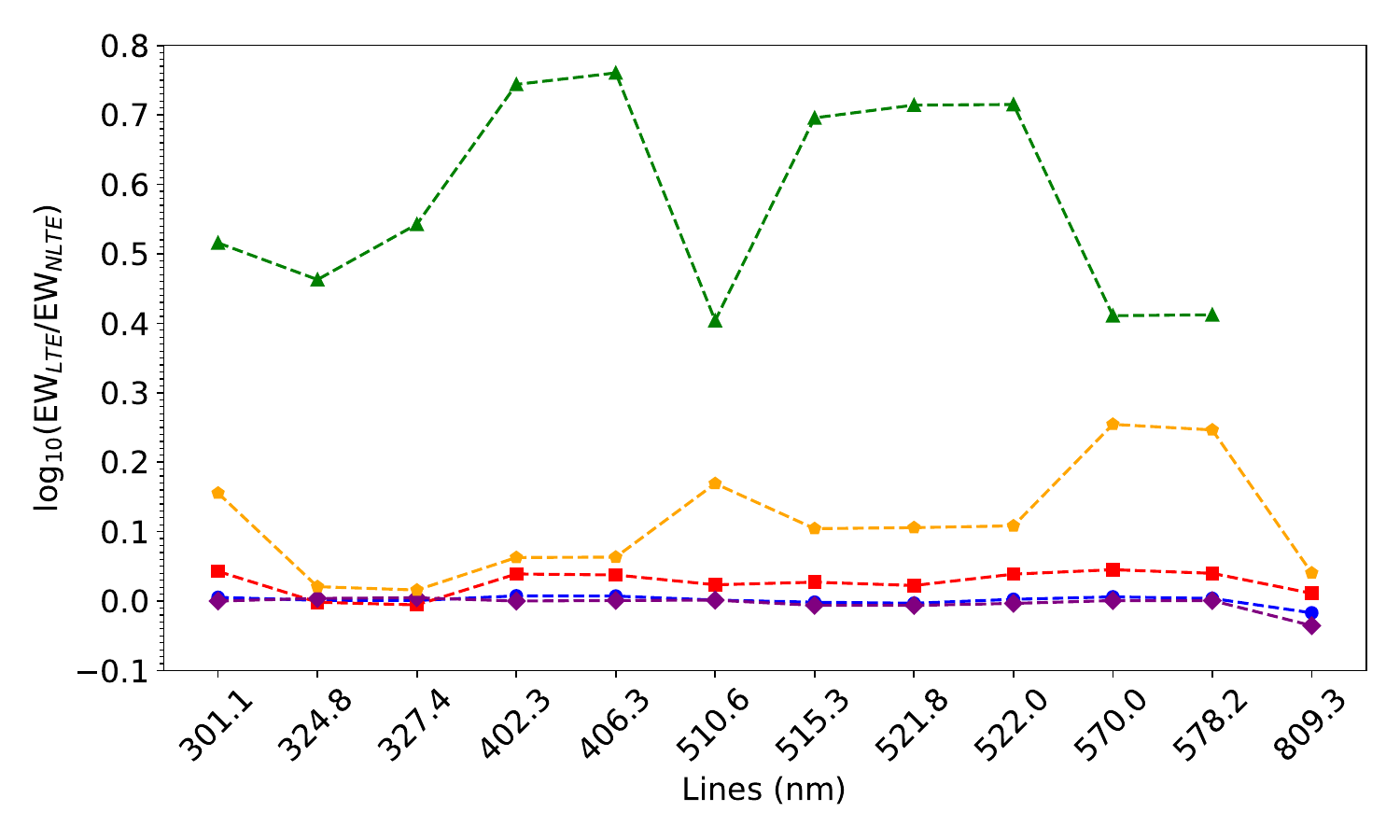}
        \includegraphics[scale=0.355]{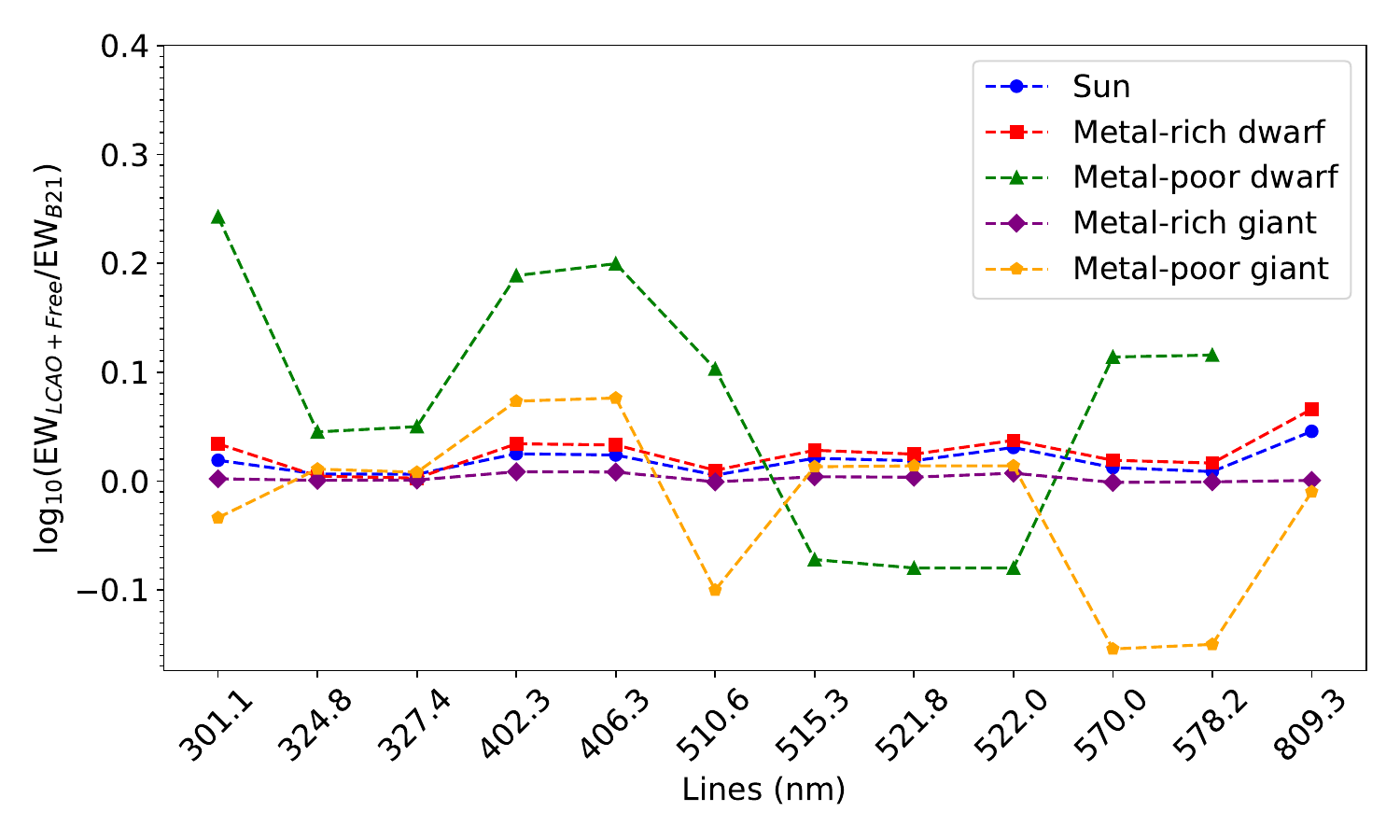}        \caption{Equivalent width ratios for the 12 \ion{Cu}{I} lines analyzed in this paper, calculated for 5 different stellar atmosphere models. The left panel is the LTE versus non-LTE EW ratios based on our recommended LCAO + Free non-LTE model. The right panel is the non-LTE with the LCAO + Free model versus non-LTE with the B21 model EW ratios. In both panels, the line at 809.3 nm for the metal-poor dwarf is not shown because the EW is too small.}
        \label{fig:LTEvsNLTE}
    \end{center}
\end{figure*}

\subsection{Cu model atom} \label{atom}

The Cu model atom used in this work is described in \cite{raccathesis}. We illustrate the Grotrian diagram of the model in \fig{fig:modelatom}. The model contains 150 energy levels, taking into account fine structure: 125 are of neutral Cu, 24 are of ionised Cu and the ground state of doubly-ionised Cu is also included. These included levels reach 0.081 eV below the first ionisation limit for \ion{Cu}{I}, and 9.2 eV below the ionisation limit for \ion{Cu}{II}. The energy levels and bound-bound radiative transitions come from the database of \cite{kurucz_2018ASPC..515...47K}.
The choice of data for photoionisation and inelastic electron collisions broadly follows that used in \cite{Andrievsky_2018MNRAS.473.3377A} and \cite{2023MNRAS.518.3796C}.
Namely, the photoionisation cross-sections are taken from R-matrix calculations by \cite{Liu_2014ApJS..211...30L}. The cross-sections for electron-impact excitation were calculated based on the semi-empirical recipe of \citet{van_regemorter_1962ApJ...136..906V}, and for electron-impact ionisation from the empirical recipe of \cite{cox_2000asqu.book.....C}.

Pressure broadening due to hydrogen collisions was included via ABO theory \citep[e.g.][]{barklem_2000A&AS..142..467B}\footnote{Interpolating extended tables at \url{https://github.com/barklem/abo-cross}.}. In addition, hyperfine structure and isotopic contributions were added for 12 \ion{Cu}{I} lines that are often discussed in the literature and that are the main focus of this study.  These values were adopted from \citet{Shi_2014ApJ...782...80S}, who considered two stable isotopes of Cu, $^{63}$Cu and $^{65}$Cu, with an isotopic fraction of 0.69 and 0.31, respectively \citep{Asplund_2009ARA&A..47..481A,Asplund_2021A&A...653A.141A}.

We then included our newly calculated rate coefficients for inelastic collisions with hydrogen, based on our asymptotic LCAO and free electron models. Different model atoms were created by only modifying the hydrogen collision data. Our recommended model is based on the combined data from the asymptotic LCAO calculation and free electron calculations (``LCAO+Free'').  To investigate the relative importance of the two sets of data, a model was created with data solely from the asymptotic LCAO calculation (hereby ``LCAO'').  However, since our model atom includes fine-structure levels, we redistributed the rate coefficients among fine-structure levels by dividing the coefficients calculated without fine-structure by the total number of final states, following Boltzmann distributions. This approach assumes that the energy differences between fine-structure levels are small enough. We also set the rates within fine-structure levels to very high values, justified by the expected efficiency of inelastic collision processes between fine-structure levels based on the Massey criterion, referred to as "relative LTE" \citep{Massey_1949RPPh...12..248M}. 

Finally, a model was created with data solely from B21 \citep{Belyaev_2021MNRAS.501.4968B}, in order to analyse the impact of hydrogen collisions on non-LTE spectra of \ion{Cu}{I} in \sect{non-LTE}. For consistency, the collapsed B21 rates were also redistributed among fine-structure levels using the same method as for the LCAO rates. Such as for the LCAO and free-electron rate, we assume infinite rates within fine-structure levels (i.e. relative LTE). We argue that for fine-structure levels that have a small energy difference, relative LTE between these levels leads to more physically motivated populations of these levels. For instance, if we consider the transition between $3d^{10} 4p$ $^2\mathrm{P}_{1/2}$ and $^2\mathrm{P}_{3/2}$, the collisional rate given in B21 is $10^{-13}$ cm$^3$/s, which is very small. Since the electron collision rates for transitions within fine-structure levels are not included, this can lead to the total collision rate for that transition being underestimated. The impact of collapsing and redistributing the B21 rate coefficients instead of using fine-structure rates on non-LTE spectra will be discussed in \sect{non-LTE}.

\subsection{Non-LTE effects across stellar parameters} \label{resultsnon-LTE}

\begin{table}
\caption{The stellar parameters of the \marcs{} model atmospheres used for the Sun, metal-rich dwarf (MRD), metal-poor dwarf (MPD), metal-rich giant (MRG), and metal-poor giant (MPG) analysed in \sect{resultsnon-LTE} and \sect{non-LTE}.}
\label{table:parameters}
\centering
\begin{tabular}{c c c c c c}
\hline\hline
 & Sun & MRD & MPD & MRG& MPG \\
\hline 
    $\teff \, (\kelvin)$ &  $5777$ &  $6500$   & $6500$ & $4500$ & $4500$ \\ 
    $\lgg$ & $4.44$ & $4.0$      & $4.0$ & $1.0$ & $1.0$ \\
    $\vmic \, (\kms)$ & $1.0$  & $1.0$ & $1.0$ & $2.0$ & $2.0$ \\
    $\feh$ & $0.0$ & $0.0$ & $-2.0$ & $0.0$& $-2.0$ \\
    $[\mathrm{Cu}/\mathrm{Fe}]$ & $0.0$ & $0.0$  & $-0.8$ & $0.0$  & $-0.4$  \\
\hline
\end{tabular}
\tablefoot{Solar reference abundances from \citet{Asplund_2021A&A...653A.141A}.}
\end{table}

To investigate the importance of non-LTE effects for stars with different parameters and understand which stars are most affected, we analysed 12 \ion{Cu}{I} lines ranging from UV to near-infrared, highlighted in blue in \fig{fig:modelatom} and discussed in \cite{Shi_2014ApJ...782...80S}. This set includes the two resonance lines at 324.8 and 327.4 nm that are important for abundance diagnostics in metal-poor dwarfs \citep[e.g.][]{2018MNRAS.480..965K,2023ApJ...953...31S}, and the strong optical lines at 510.6 and 578.2 nm, previously used for observations of Cu abundances in metal-poor giants \citep[e.g.][]{Ishigaki_2013ApJ...771...67I, Lombardo_2022A&A...665A..10L, 2023MNRAS.518.3796C}.

The 1D non-LTE spectra were obtained using the 3D non-LTE radiative transfer code \balder{} \citep{Amarsi_2018A&A...615A.139A}, which is a custom version of \multitd{} \citep{botnen_1999ASSL..240..379B, Leenaarts_2009ASPC..415...87L}. For these test calculations, we took five \marcs{} model atmospheres across the HR diagram.  
These include a plane-parallel ``dwarf'' and a spherical ``giant'' with solar mass.  In both cases, the analysis was performed on both ``metal-rich'' and ``metal-poor'' models.  We also analysed the Sun using the standard \marcs{} solar model. The stellar parameters of the model atmospheres are described in \tab{table:parameters}.

To estimate the departure from LTE for each star and each line, we plotted the equivalent width (EW) ratios in LTE versus non-LTE in \fig{fig:LTEvsNLTE}. These EW ratios provide a direct estimate of the abundance corrections for each line, since we are in the linear regime of the curve of growth. However, one should be careful with the lines at 324.8 and 327.4 nm, as we found that these lines are saturated for the metal-poor giant. The non-LTE model is based on our recommended hydrogen collisions recipe, the combined LCAO+Free rates.

Based on the EW ratios, we can see that the non-LTE effects for \ion{Cu}{I} are strongest in the metal-poor regime, which was confirmed already by \cite{Shi_2018ApJ...862...71S} and \cite{Andrievsky_2018MNRAS.473.3377A}. The effects are the strongest for the hot metal-poor dwarf, with an average correction of +0.6 dex. However, for some of the largest EW ratios, one should be careful since some of those lines are extremely weak, with a logarithmic reduced EW $<-$ 8 (e.g. the lines at 402.3 and 406.3 nm).  Nevertheless, the two resonance lines give a correction of $\sim$ +0.5 dex for the given $[\mathrm{Cu}/\mathrm{Fe}]$, which is important to take into account when observing these lines for abundance diagnostics. The metal-poor giant also shows important non-LTE corrections, with up to +0.25 dex for the broadly used optical lines at 570.0 and 578.2 nm. The corrections are more severe for stars of higher temperature and lower metallicity, which is likely linked to the stronger UV radiation field driving a stronger overionization of \ion{Cu}{I}, which also explains the differences between the hotter dwarf and cooler giant. Importantly, one should note that the size of the non-LTE corrections is heavily influenced by the Cu abundance for all lines due to competing non-LTE effects. The non-LTE corrections showed in \fig{fig:LTEvsNLTE} were calculated considering typical Cu abundances in metal-poor dwarfs and giants found in stars described in \sect{gce} (see $[\mathrm{Cu}/\mathrm{Fe}]$ in \tab{table:parameters}).

Overall, this test shows us that the line-averaged non-LTE abundance corrections are positive in four of the five stars, with only the metal-rich giant being very slightly negative. This means that Cu abundances tend to be pushed towards higher values in non-LTE, which is in agreement with previous studies of \ion{Cu}{I}. A more detailed calculation and analysis of non-LTE abundance corrections for observed stars in the literature will be done in \sect{gce}.

\subsection{Sensitivity to the inelastic Cu + H collisions} \label{non-LTE}

To test the effect of hydrogen collision data on the non-LTE line profiles and abundance corrections, we computed the same 12 lines as previously mentioned using two different hydrogen data sets: the data from our LCAO+Free calculations, and the data from B21. On the right panel of \fig{fig:LTEvsNLTE}, we compare the equivalent widths using the two data sets for the 5 different stellar atmosphere models that we tested. This allows us to do a differential analysis by comparing the effect of hydrogen collision data from us and from B21 on non-LTE spectra. Moreover, it allows us to see for which stellar parameters and for which \ion{Cu}{I} lines we expect a larger sensitivity to the Cu+H recipe, and thus to quantify the uncertainty on the hydrogen collision (within the NLTE modelling uncertainty). 

As expected, we can see that in the metal-poor regime, the differences in EW ratios can be non-negligible. Although metal-poor giants are expected to show the greatest sensitivity to hydrogen collisions due to their lower atmospheric pressure and cooler temperatures compared to metal-poor turn-off stars, the largest absolute EW ratios are observed in metal-poor dwarfs, reaching up to $\sim$0.3 dex. However, this is partly because metal-poor dwarfs exhibit very large non-LTE effects, averaging around 0.6 dex, while the effect of hydrogen collisions is $\sim$ 0.1 dex on average. In contrast, for metal-poor giants, the average non-LTE correction is smaller, around 0.1 dex, while the impact of hydrogen collisions using our new data is 0.05 dex (on average). This means that, relative to the magnitude of the abundance correction, metal-poor giants exhibit the highest sensitivity to hydrogen collisions, while the large non-LTE effects in metal-poor dwarfs diminish the relative impact of hydrogen collisions in those stars.

\begin{figure*}[htbp]
    \centering
    \includegraphics[scale=0.21]{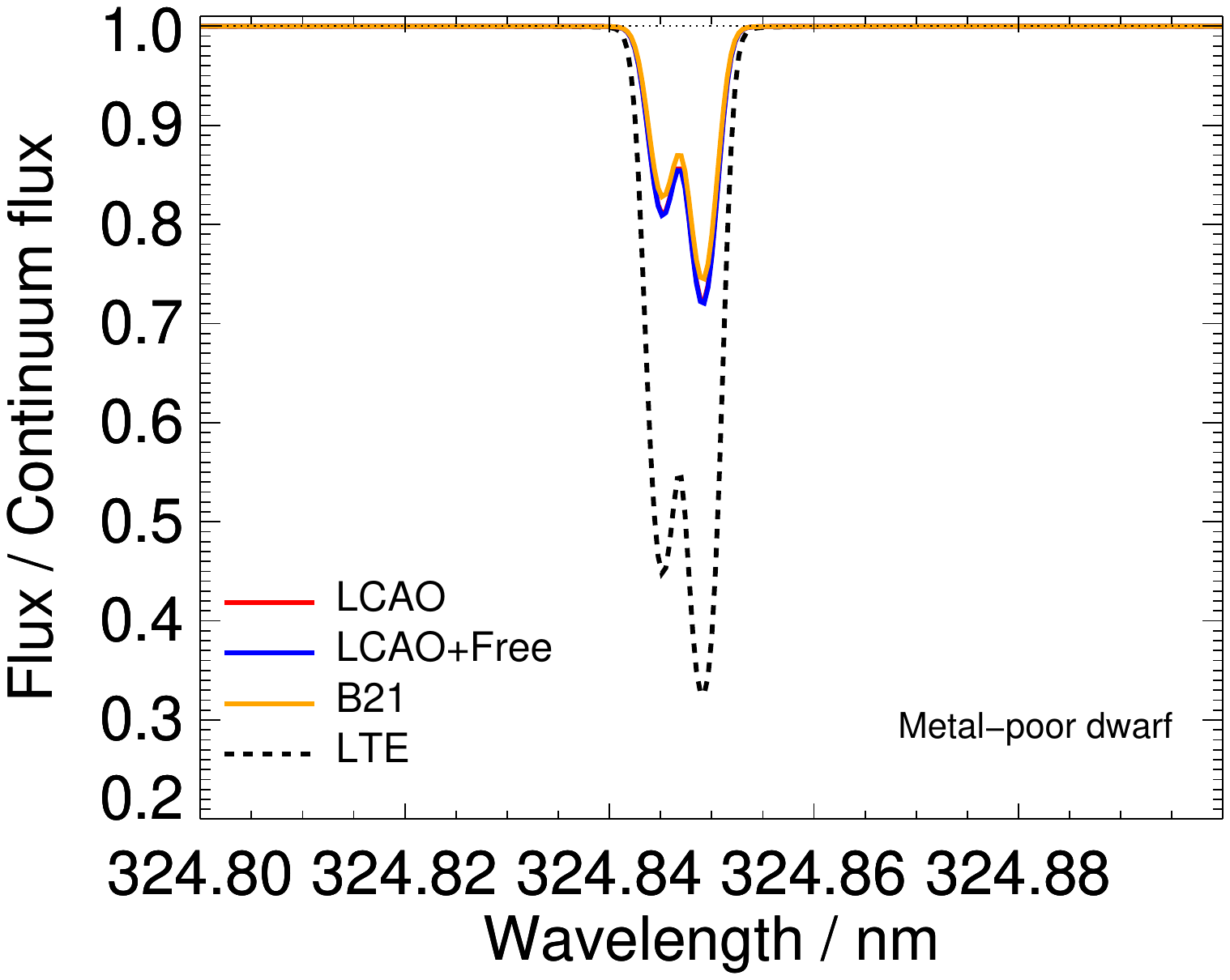}\includegraphics[scale=0.21]{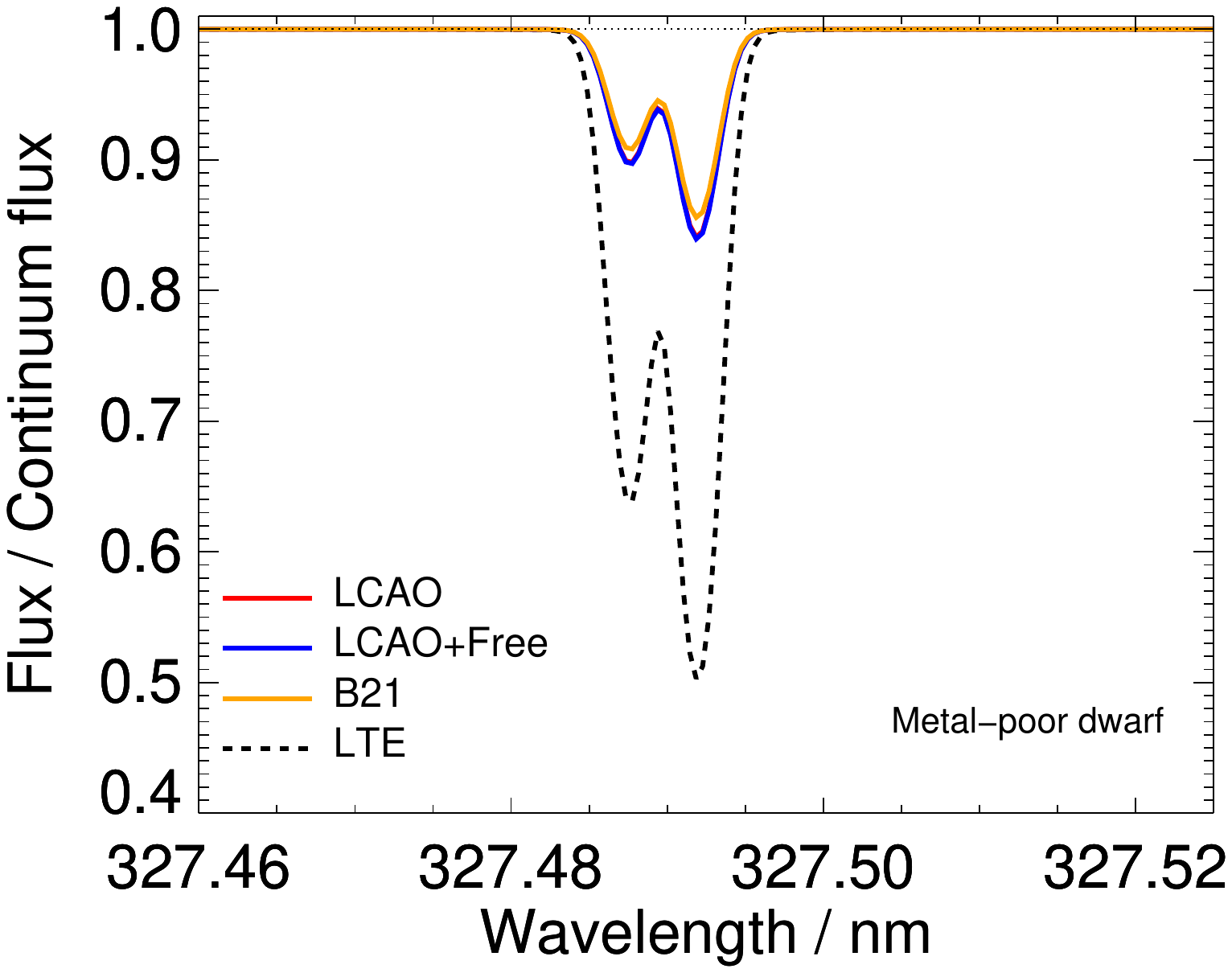}\includegraphics[scale=0.21]{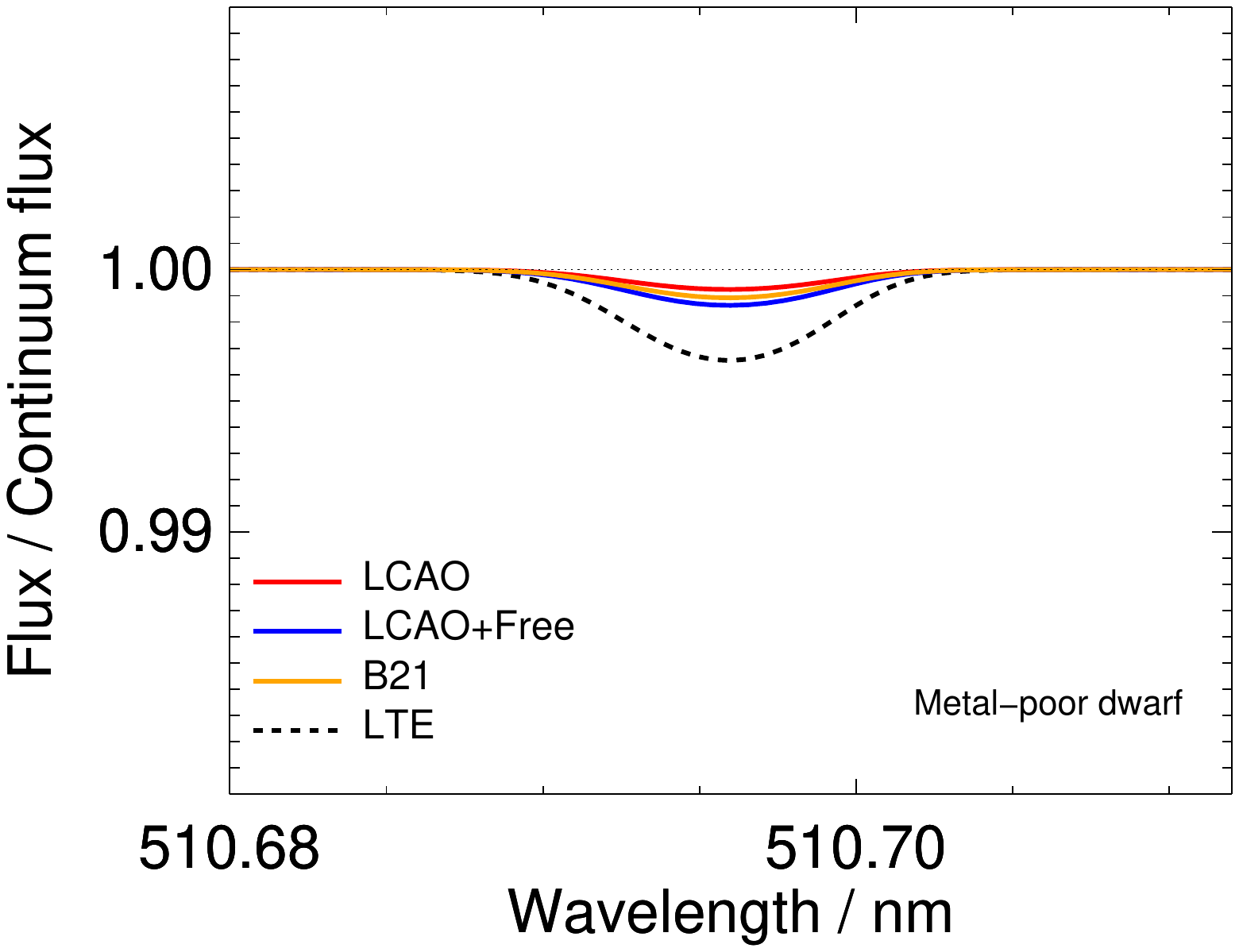}
    \includegraphics[scale=0.21]{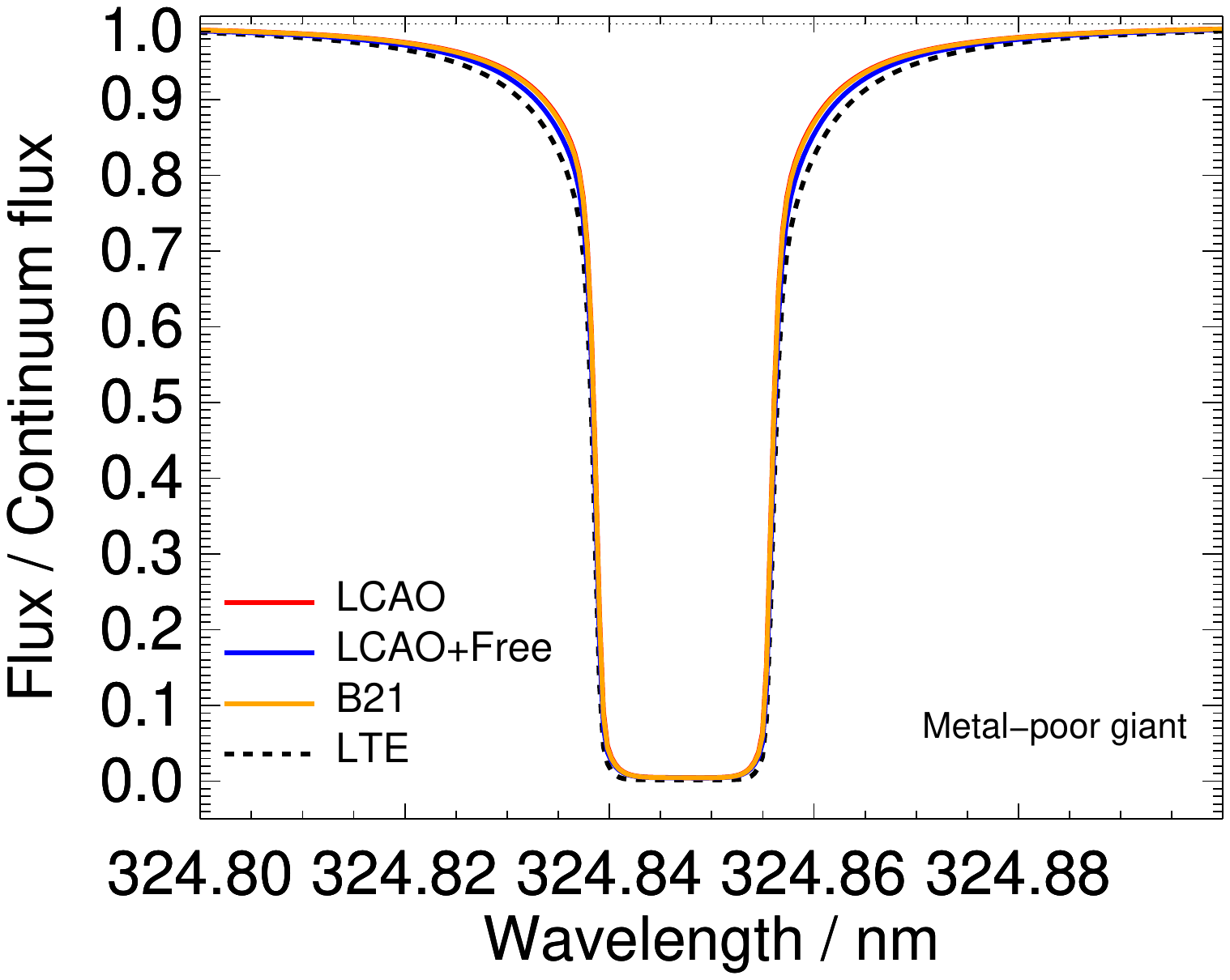}\includegraphics[scale=0.21]{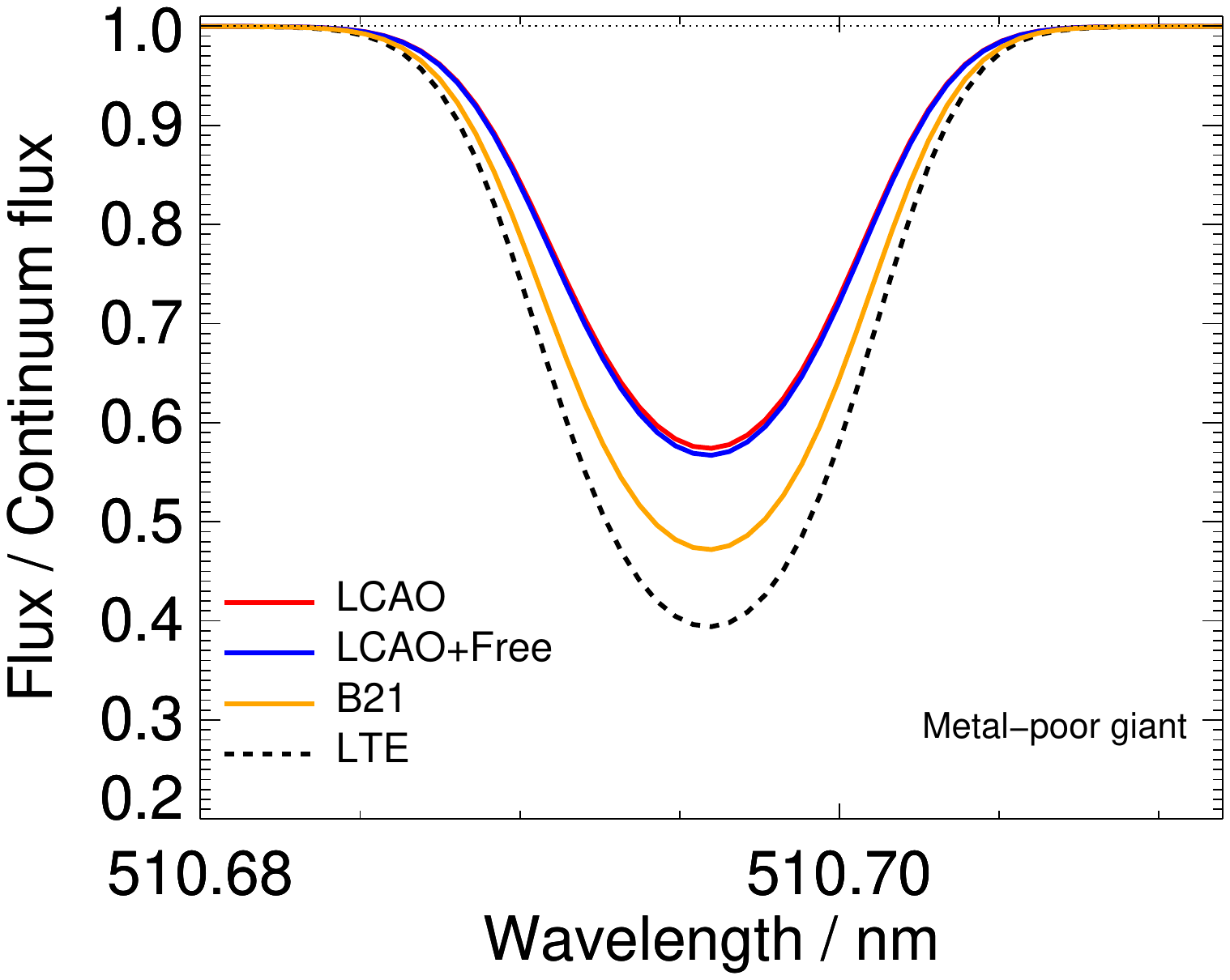}\includegraphics[scale=0.21]{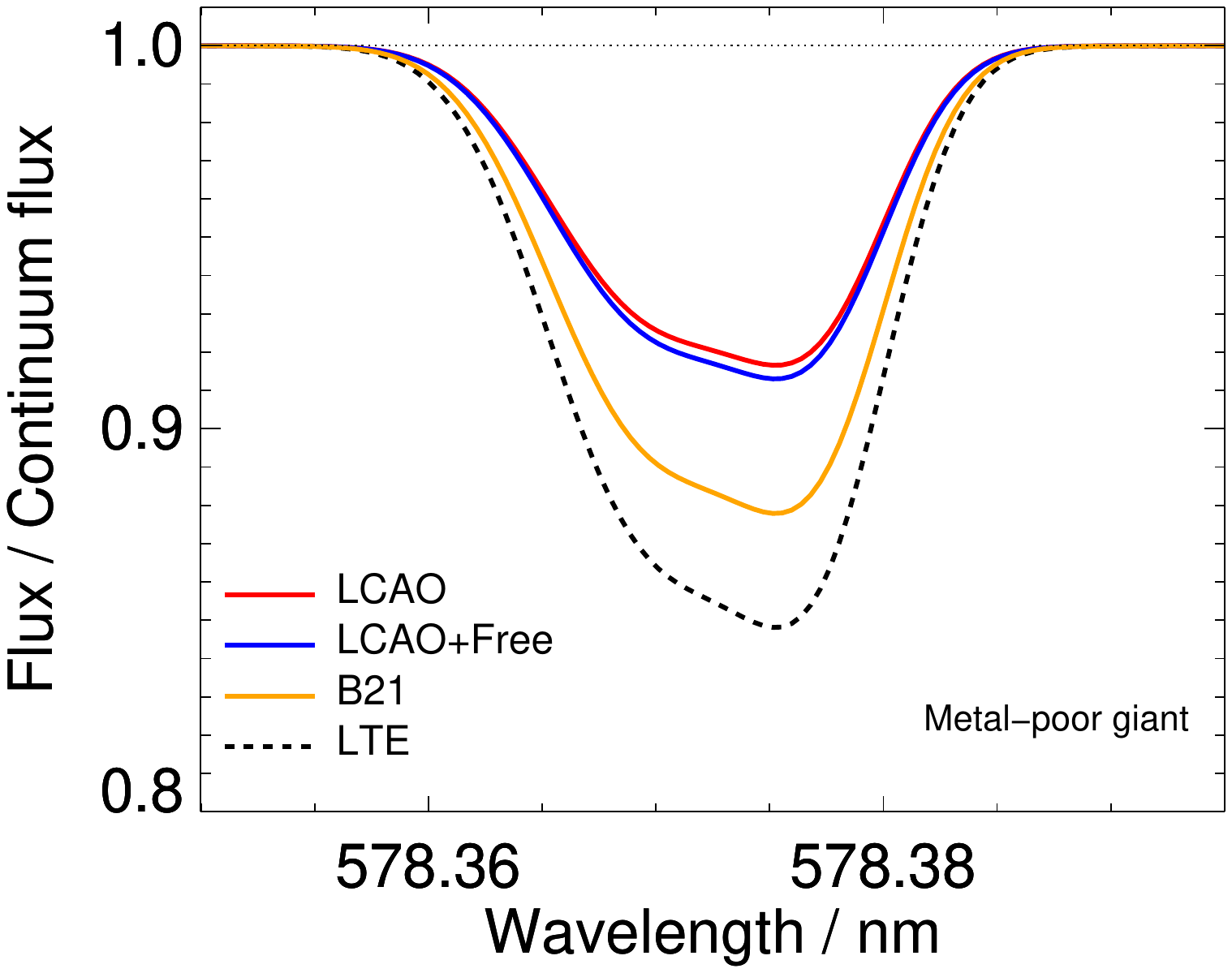}
    \caption{Comparison of LTE (dashed black line) and non-LTE (solid lines) modelling for the 324.8, 327.4, 510.6 and 578.2 nm \ion{Cu}{I} lines in a metal-poor dwarf (top 4 panels) and a metal-poor giant (bottom 4 panels). The impact of different Cu+H collision data on the non-LTE profiles is shown, with red and blue lines using newly calculated rates, while the orange line represents rates from \citet{Belyaev_2021MNRAS.501.4968B}. No macroscopic broadening (macroturbulence, instrumental, and rotational broadening) was added to the lines.}
    \label{fig:lines}
\end{figure*}

We can also discuss which lines are more sensitive to hydrogen collision data. This can be important to take into account when doing abundance diagnostics. In \fig{fig:lines}, we plotted some of the interesting non-LTE line profiles for the metal-poor dwarf and giant (since these stars are where we see the biggest sensitivity), based on different Cu+H collision recipes. First of all, we can see that the two most used resonance lines are quite sensitive to the hydrogen collision data for the dwarf, but they are saturated in the giant (some variation can be seen in the wings). The two optical lines at 510.6 and 578.2 nm are also part of the lines that show a large sensitivity, especially in the metal-poor giant. Moreover, the line profiles show that adding the rates from the free electron model to our asymptotic LCAO rates actually has an impact on the line profiles, especially for the metal-poor dwarf. In particular, the 510.6 nm line seems to be very sensitive to the rates from the free electron model for the metal-poor dwarf, while it is much less sensitive in the giant. Finally, we can see from \fig{fig:LTEvsNLTE} that lines belonging to the same multiplet (i.e. originating from the fine-structure levels of the same states) such as the lines at 510.6, 570.0 and 578.2 nm, exhibit similar LTE versus non-LTE EW ratios, as well similar sensitivity to the hydrogen collision data. 

To understand the cause of the differences between the non-LTE line profiles modelled using the rates from B21 and the rates from the LCAO model, we did different tests. First, we verified that we did not introduce any errors in the B21 data during the collapsing and redistribution of the rate coefficients among fine-structure levels. Indeed, since the rates from B21 account for fine-structure while the rates from the LCAO + Free model do not, in order to compare the rates, we collapsed the Belyaev rates and redistributed them the same way as the LCAO rates. We checked the impact of the collapsing and redistributing the rates by comparing the equivalent widths of the non-LTE line profiles modelled using the original B21 rates with fine-structure and in relative LTE and the redistributed B21 rates. For the Sun, the redistributed rates do not influence the line profiles. The difference between the lines increases for the metal-poor regime, but even then it averages around $\sim$ 0.003 dex, which is negligible within the uncertainties on the hydrogen collision data.

Then we carried out tests to determine which rate coefficients are responsible for the differences in non-LTE line profiles, by switching off selective rate coefficients in the model atom. We found that for the metal-poor stars, the biggest differences arise from the de-excitation rates for the $ 3d^{10}4p \, ^2\mathrm{P} \rightarrow 3d^{9}4s^2 \, ^2\mathrm{D}$ transitions. As explained in \sect{methodLCAO}, these transitions give rate coefficients that are zero in our LCAO model. However, they are among the strongest rates in B21, being on the order of charge transfer rates ($\sim$ 10$^{-10}$). As we can see from \fig{fig:modelatom}, the optical lines at 510.6, 570.0 and 578.2 nm are from transitions between the $3d^{9}4s^2 \, ^2\mathrm{D}$ and $3d^{10}4p \, ^2\mathrm{P}$ states, so it makes sense that these collisional rates have a big impact on those lines.

\section{Astrophysical validation and implications on the Galactic evolution of copper}\label{gce}

In order to see the impact of our newly calculated hydrogen collisions data, we applied our 1D non-LTE corrections to 1D LTE Cu abundances of stars observed in the literature.

\subsection{Calculation of grids of abundance corrections}
 
Calculations were performed on two subsets of the \marcs{} grid of model atmospheres \citep{2008A&A...486..951G}.  The first set covers dwarfs and subgiants with $\teff$ from $5000\,\kelvin$ to $7000\,\kelvin$ and $\lgg$ from $3.5 \dex$ to $5.0\,\dex$; while the second set covers giants with $\teff$ from $4000\,\kelvin$ to $5000\,\kelvin$ and $\lgg$ from $0.5 \dex$ to $2.5\,\dex$.  Both sets cover the same range of metallicities, with $\feh$ from $-4.0$ to $+0.5$.  These were drawn from the grid employed in \citet{2020A&A...642A..62A}. As in \sect{methodnon-LTE}, \balder{} was used to calculate 1D LTE and 1D non-LTE equivalent widths and abundance corrections for these models.  The calculations adopted a range of abundances with $\xfe{Cu}$ between $-1.2\,\dex$ and $+1.2\,\dex$ in steps of $0.4\,\dex$, and $\vmic$ from $0\,\kms$ to $3\,\kms$ in steps of $1\,\kms$.

\subsection{Literature data}\label{litdata}

The abundance corrections were applied to 1D LTE results from the literature, based on high resolution spectroscopy and spanning a wide range of metallicities. At the most metal-rich end, 1D LTE data from \citet{2017A&A...606A..94D} were adopted. The authors provide copper abundances for $1058$ FGK-type dwarfs and subgiants. Since the authors do not provide line-by-line abundances, we restricted the sample to those for which the final abundance was based on all four lines.  We also limited it to stars within our grid of calculations, and to stars not present in the \citet{Yan_2016A&A...585A.102Y} data set described below.  The final count of stars from this sample becomes $698$. Their stellar parameters were derived in previous works from the same series (see references therein) based on excitation and ionisation equilibrium of \ion{Fe}{I} and \ion{Fe}{II} lines in 1D LTE; at these metallicities, the errors incurred in the stellar parameters by assuming 1D LTE are smaller and so we do not attempt to correct for them here.   The authors measured copper abundances using the \ion{Cu}{I} $510.6\,\nm$, $521.8\,\nm$, $522.0\,\nm$, and $578.2\,\nm$ lines. They adopted oscillator strengths from \citet{1975JQSRT..15..463B}, with $\log gf$ values of $-1.52\,\dex$, $+0.48\,\dex$, $-0.45\,\dex$, and $-1.72\,\dex$ respectively.  For the \ion{Cu}{I} $521.8\,\nm$ and $522.0\,\nm$ lines, these oscillator strengths are around $0.2\,\dex$ larger than those of \citet{1968ZA.....69..180K}; nevertheless, these uncertainties are reduced here because their analysis is differential with respect to the Sun and by averaging over four lines. We calculated and applied line-averaged solar-differential 1D non-LTE corrections to them \footnote{[Cu/H]$_{\mathrm{NLTE}}$ $=$ [Cu/H]$_{\mathrm{LTE}}$ + ($\Delta_{\mathrm{star}} - \Delta_{\mathrm{sun}} $)}.

Going down in metallicity, the line-by-line 1D LTE copper abundances from \citet{Yan_2016A&A...585A.102Y} were adopted.  This sample covers $78$ FG-type dwarfs and subgiants.  The data set comprises observations with two instruments (NOT/FIES and VLT/UVES) and we used the average result for the few stars measured by both instruments.  Their stellar parameters are based on the ``inverted'' spectroscopic analysis described in \citet{2010A&A...511L..10N}. The iron abundances from \citet{Yan_2016A&A...585A.102Y}, based on \ion{Fe}{II} lines in 1D LTE, are close to the 3D LTE ones derived in \citet{2019A&A...630A.104A} and adopted in \citet{2024A&A...682A.116N} for the $73$ stars in common.
Their copper abundances were derived from the \ion{Cu}{I} $510.6\,\nm$, $521.8\,\nm$, and $578.2\,\nm$ lines.  They derived solar-calibrated oscillator strengths although their values are unspecified; the text suggests that this was under the assumption of $\lgeps{Cu}=4.25$ that they adopted from \citet{2009LanB...4B..712L}.  Here, we took their absolute abundances, applied absolute 1D non-LTE corrections to these results on a line-by-line basis, and then converted to a line-averaged $\xfe{Cu}$ using the solar abundance of $\lgeps{Cu}=4.18$ from \citet{Asplund_2021A&A...653A.141A}. 

Next, we adopted line-by-line 1D LTE abundances from \citet{2018MNRAS.480..965K}.  This sample contains six metal-poor dwarfs and subgiants, of which we adopt five (HD76932 being present in the data set of \citealt{Yan_2016A&A...585A.102Y}, described above).  Their stellar parameters are from the earlier study of \citet{Roederer_2018ApJ...857....2R}, based on photometry and parallaxes.  We took the mean of the two values of $\feh$ that are given in Table 3 by \citet{Roederer_2018ApJ...857....2R}, determined from \ion{Fe}{I} lines and from \ion{Fe}{II} lines respectively.  We took 1D LTE absolute copper abundances based on an average of \ion{Cu}{I} $324.8\,\nm$ and $327.4\,\nm$ resonance lines from Table 6 and Figure 3 of \citet{2018MNRAS.480..965K}; these lines have well-constrained oscillator strengths as discussed in \citet{Roederer_2018ApJ...857....2R}.  We applied absolute line-averaged 1D non-LTE corrections to them: the corrections are similar for the two lines, being within $0.05\,\dex$ for all five stars.  We then used the solar abundance $\lgeps{Cu}=4.18$ from \citet{Asplund_2021A&A...653A.141A} to convert to $\xfe{Cu}$.

At the lowest metallicities, 1D LTE abundances were taken from \citet{2023ApJ...953...31S}.  They presented copper abundances in $36$ FG-type dwarfs and subgiants, and we remove one star from the sample that falls outside of our grid of abundance corrections (being too low in surface gravity).  The stellar parameters are based on a spectroscopic analysis in 1D LTE (excitation balance of \ion{Fe}{I}, and ionisation balance of several elements including iron. 
Here, as above, we simply take the mean of the \ion{Fe}{I} and \ion{Fe}{II} results to obtain [Fe/H]. We took their line-averaged 1D LTE copper abundances (converting $\xfe{Cu}$ from their Table 2 to the absolute scale, consistently with their analysis), that are based on the \ion{Cu}{I} $324.8\,\nm$ and $327.4\,\nm$ resonance lines. As above, we applied line-averaged absolute 1D non-LTE corrections to them: in this case the median difference between corrections for the two lines is just $0.02\,\dex$ and the largest difference is less than $0.04\,\dex$.  We then used the solar abundance $\lgeps{Cu}=4.18$ from \citet{Asplund_2021A&A...653A.141A} to convert to $\xfe{Cu}$.

Finally, we considered the metal-poor giants analysed by \citet{2023MNRAS.518.3796C}.  We took a subset of their data, namely the ones having copper measurements based on observations with OHP/SOPHIE (generally of higher signal-to-noise than the TBL/Neo-Narval sample), and limited to stars within our grid ($\lgg\geq0.5$).  This left us with $18$ stars.  Their effective temperatures and surface gravities are based on \textit{Gaia} photometry and parallaxes, following a procedure described in \cite{2021A&A...656A.155L}.  The authors report small non-LTE effects for their \ion{Fe}{I} lines (at most $0.03\,\dex$) as well as good agreement overall between \ion{Fe}{I} and \ion{Fe}{II} lines.  Most of their stars fall outside of the 3D LTE grid of \ion{Fe}{II} lines in \citet{2019A&A...630A.104A}, which stops at $\lgg=1.5$.  Nevertheless, testing at the boundaries of that grid suggests small 3D corrections that are at most of the order $0.05\,\dex$.  Their copper abundances are based on the \ion{Cu}{I} $510.6\,\nm$, $521.8\,\nm$, $570.0\,\nm$, and $578.2\,\nm$ lines with experimental oscillator strengths from \citet{1968ZA.....69..180K} with $\log gf$ values of $-1.50\,\dex$, $+0.26\,\dex$, $-2.33\,\dex$, and $-1.78\,\dex$ respectively.  As pointed out above, there are significant uncertainties in the oscillator strengths for optical \ion{Cu}{I} lines; for example the $\log gf$ for the \ion{Cu}{I} $521.8\,\nm$ is $0.2\,\dex$ larger in the dataset of \citet{1975JQSRT..15..463B}.  Here, for a given star, the copper abundance is based on the average result of at least two lines.  We took their absolute 1D LTE abundances and applied 1D non-LTE corrections to them on a line-by-line basis, and used the solar abundance $\lgeps{Cu}=4.18$ from \citet{Asplund_2021A&A...653A.141A} to derive a line-averaged $\xfe{Cu}$.  

\subsection{Line-to-line scatter in metal-poor giants}

\begin{figure*}[ht]
    \begin{center}
        \includegraphics[scale=0.32]{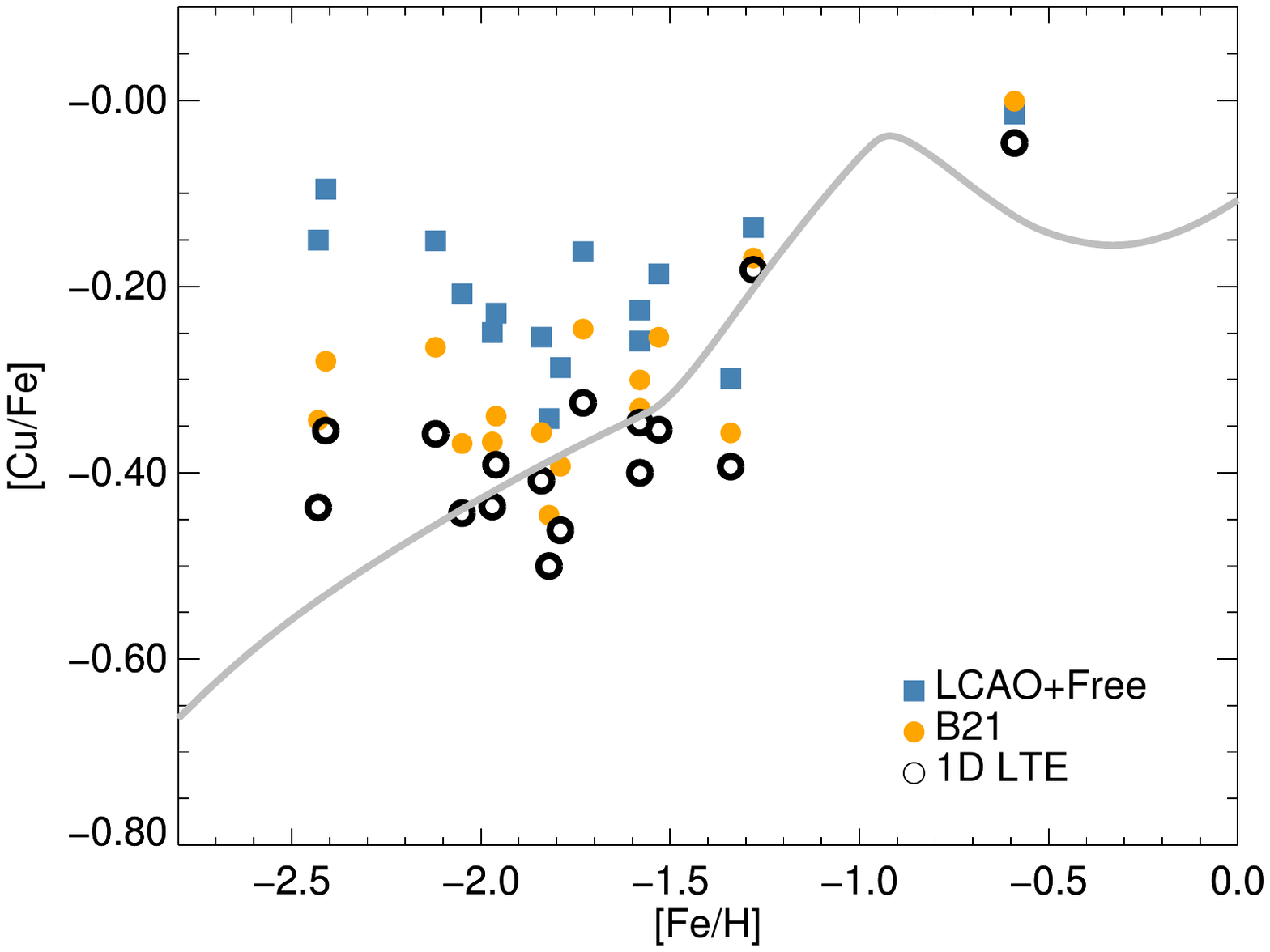}
        \includegraphics[scale=0.32]{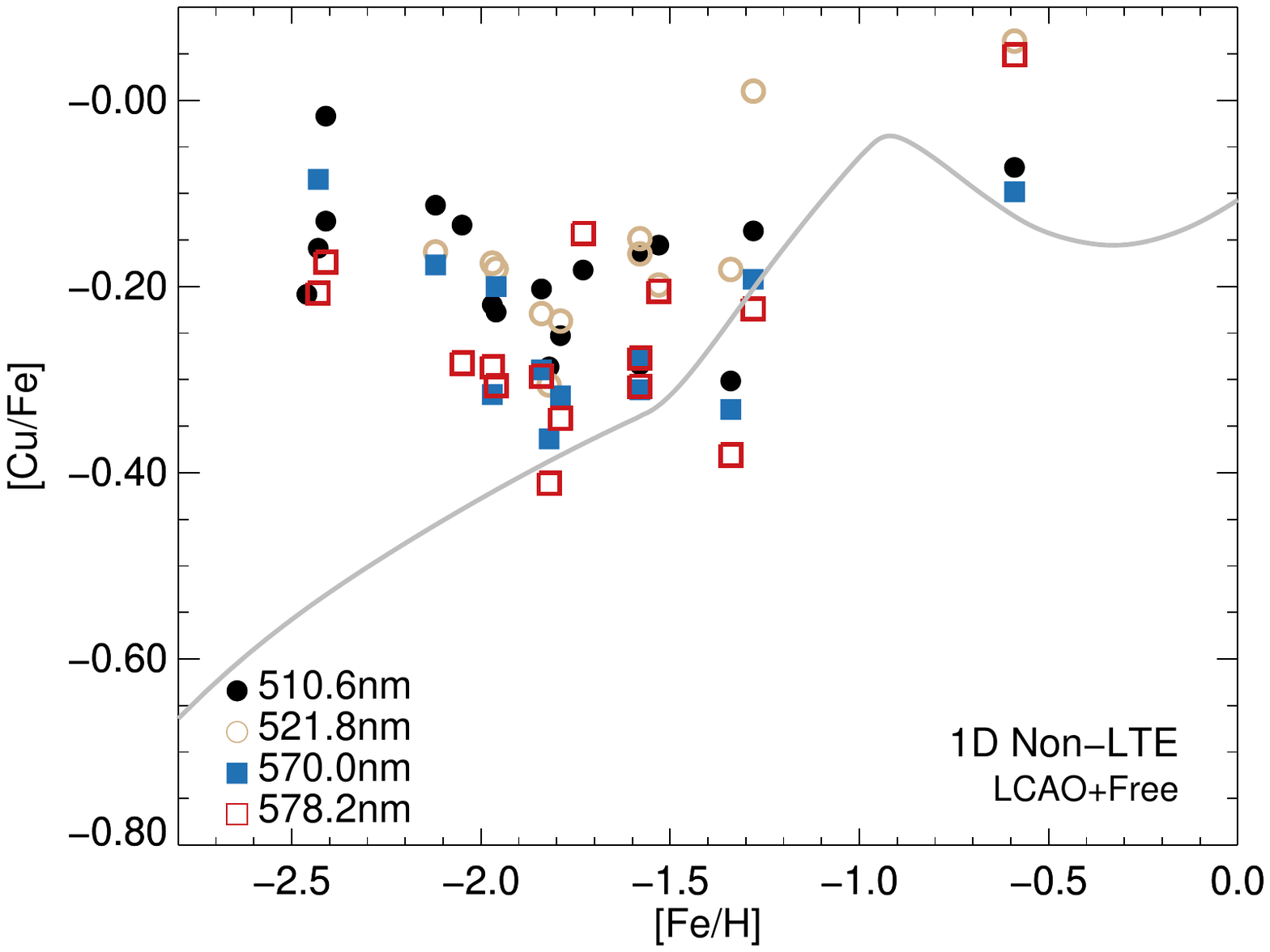}
        \caption{[Cu/Fe] versus [Fe/H] for the metal-poor giants from the \citet{2023MNRAS.518.3796C} sample, with the GCE model from \citet{2020ApJ...900..179K} overplotted in each panel. Left panel: line-averaged abundances in LTE and in non-LTE with Cu + H collision data from the present calculation (LCAO + Free) and from \citet{Belyaev_2021MNRAS.501.4968B} (B21). Right panel: line-by-line non-LTE abundances based on the Cu + H collision data from the LCAO + Free model.}
        \label{fig:copper_reanalysis_giants}
    \end{center}
\end{figure*}

\begin{figure}[ht]
\centering
\includegraphics[width=\hsize]{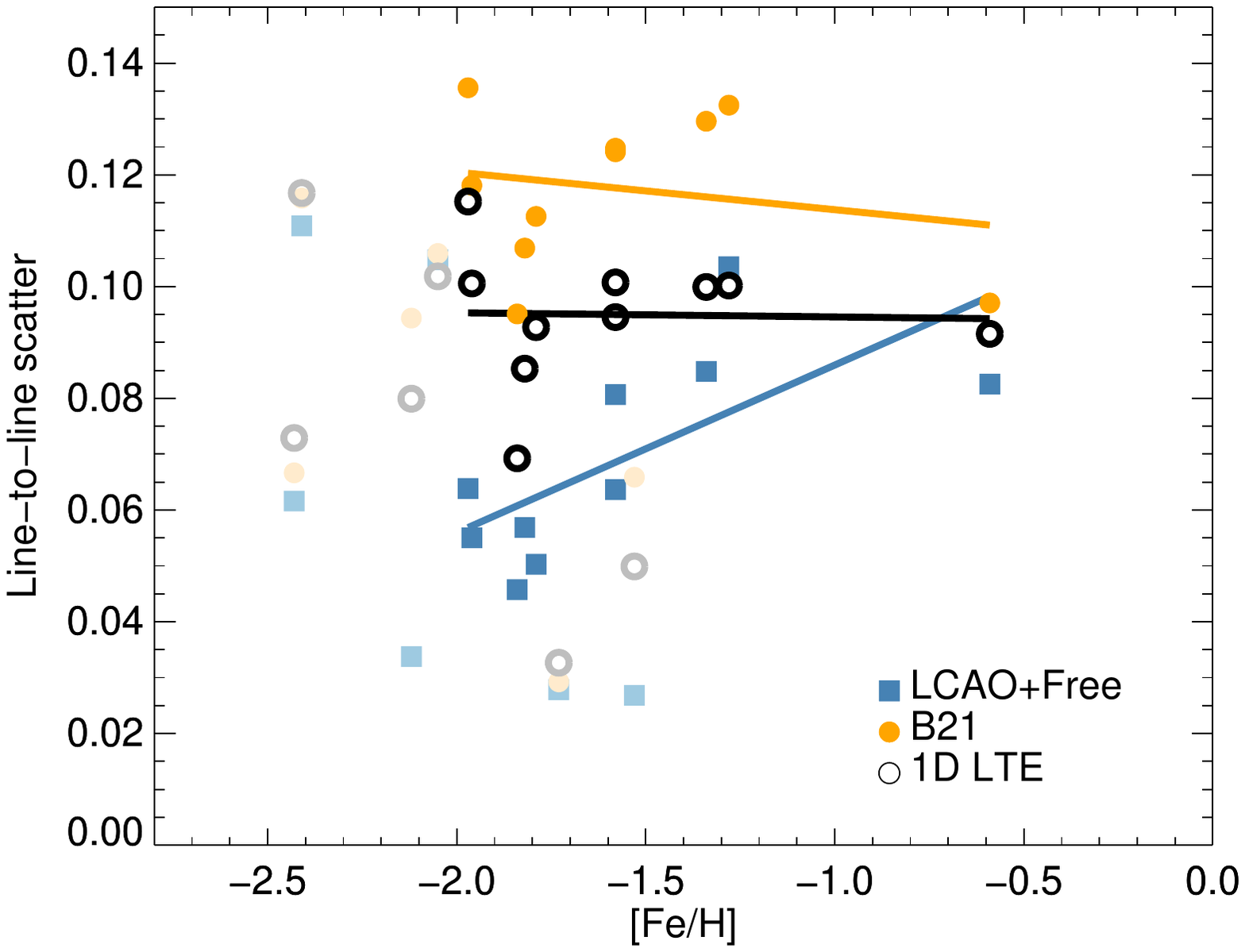}
\caption{Line-to-line scatter of the abundances in \fig{fig:copper_reanalysis_giants} (left panel). The scatter based on abundances derived from all four \ion{Cu}{I} lines as well as the fit (represented by straight lines) have dark-colored symbols, while the ones from abundances derived from only two or three lines have lighter symbols.
\label{fig:scatter}}
\end{figure}

In \sect{non-LTE}, we found that hydrogen collision data has a significant relative impact on non-LTE corrections for metal-poor giants. To investigate this further, we analyzed the metal-poor giants from the \cite{2023MNRAS.518.3796C} sample in more detail. Additionally, the availability of line-by-line abundances allows us to verify the presence of line-dependent effects on non-LTE abundance corrections and to thereby use the line-to-line scatter to validate the modelling.

In \fig{fig:copper_reanalysis_giants}, we present the non-LTE [Cu/Fe] abundances in these giants as a function of metallicity, using hydrogen collision data from both the LCAO+Free model and B21. For comparison, we also include LTE abundances and overlay a Galactic chemical evolution (GCE) model for Cu from \cite{2020ApJ...900..179K}. The non-LTE versus LTE trends look very different, with non-LTE abundances deviating more strongly from the GCE model. While the LTE trend shows a more linear metallicity dependence, we observe an ``upturn'' in [Cu/Fe] towards lower [Fe/H] $<$ $-$1.7. The choice of hydrogen collision data has a noticeable impact on the non-LTE abundance trends. The overall trends appear qualitatively different when using LCAO+Free compared to B21. Notably, the LCAO+Free model predicts stronger non-LTE effects than B21 for metal-poor giants and produces a more pronounced upturn at low metallicity. To ensure that the observed trends, particularly the low-metallicity upturn with the LCAO+Free model, are not artifacts of uncertainties in individual \ion{Cu}{I} lines, we also examined the line-by-line abundances. The different lines consistently indicate a hint of an upturn at low [Fe/H], reinforcing the robustness of this feature.

We further analysed the line-to-line scatter by calculating the sample standard deviation. As shown in \fig{fig:scatter}, the scatter is reduced in non-LTE for stars where all four \ion{Cu}{I} lines are measured (represented by darker colors). Our 1D non-LTE model incorporating the new Cu+H data (LCAO+Free) results in lower scatter compared to both LTE and the same non-LTE model using older Cu+H data (B21), increasing confidence in our approach. At higher metallicities, the scatter approaches LTE values. However, since non-LTE effects are weaker in this regime, the scatter may instead be dominated by uncertainties in the oscillator strengths. Additionally, measurement-related issues could contribute, as the \ion{Cu}{I} lines become stronger in metal-rich stars, potentially leading to saturation effects. The presence of strong hyperfine structure may further complicate the fitting of these lines.

\subsection{Galactic evolution of copper from dwarfs and giants}

\begin{figure*}[ht!]
\centering
        \includegraphics[scale=0.2175]{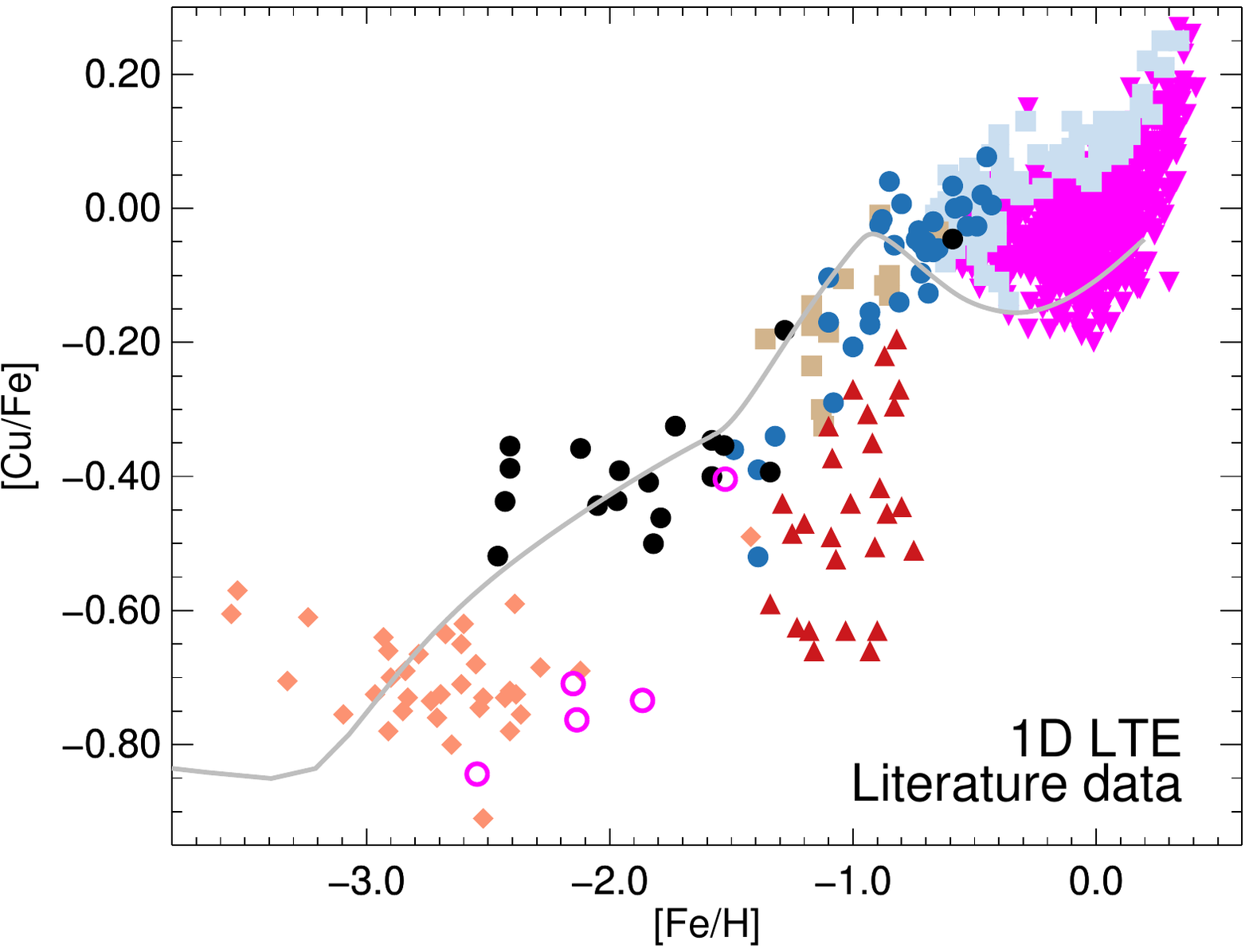}
                \includegraphics[scale=0.2175]{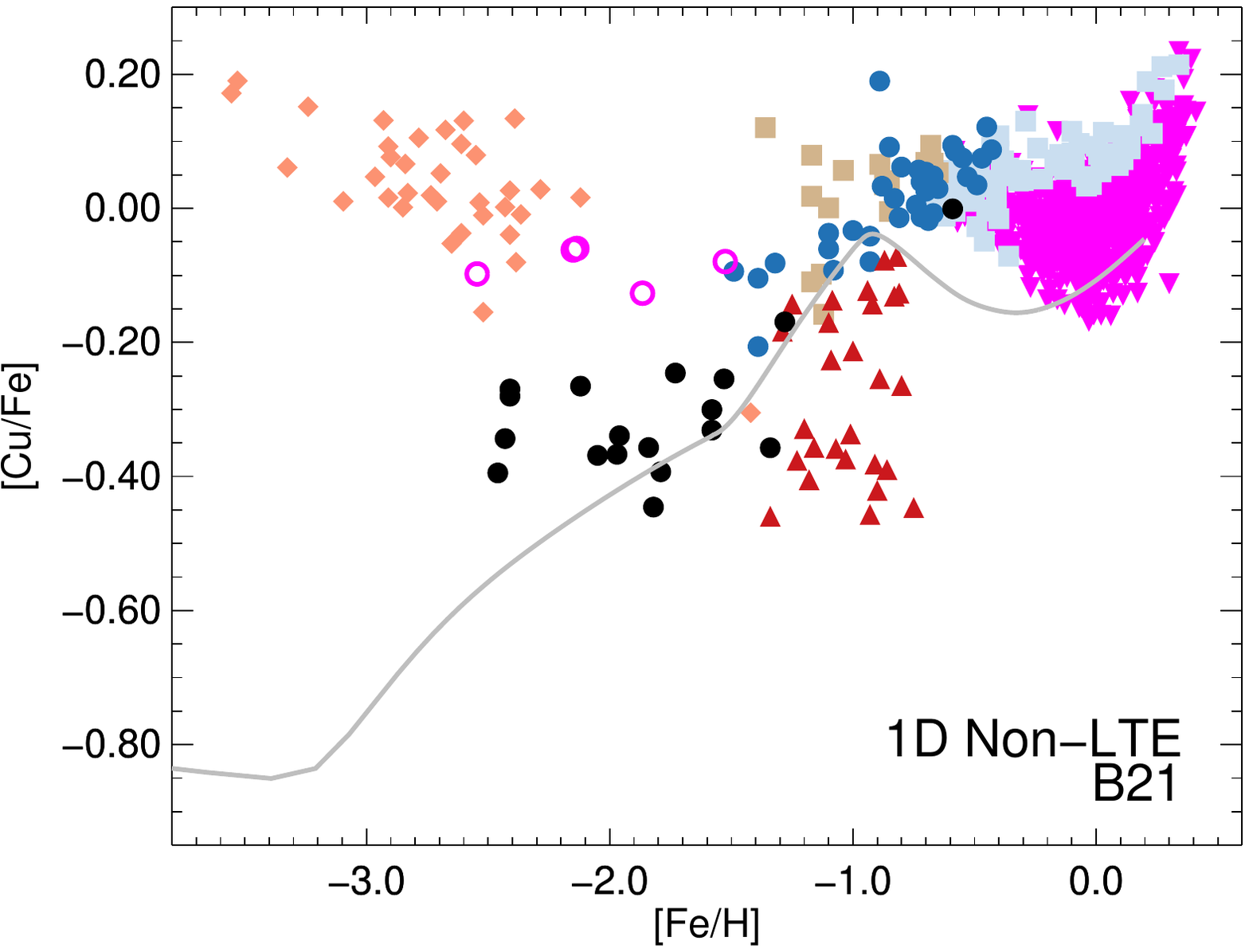}
        \includegraphics[scale=0.2175]{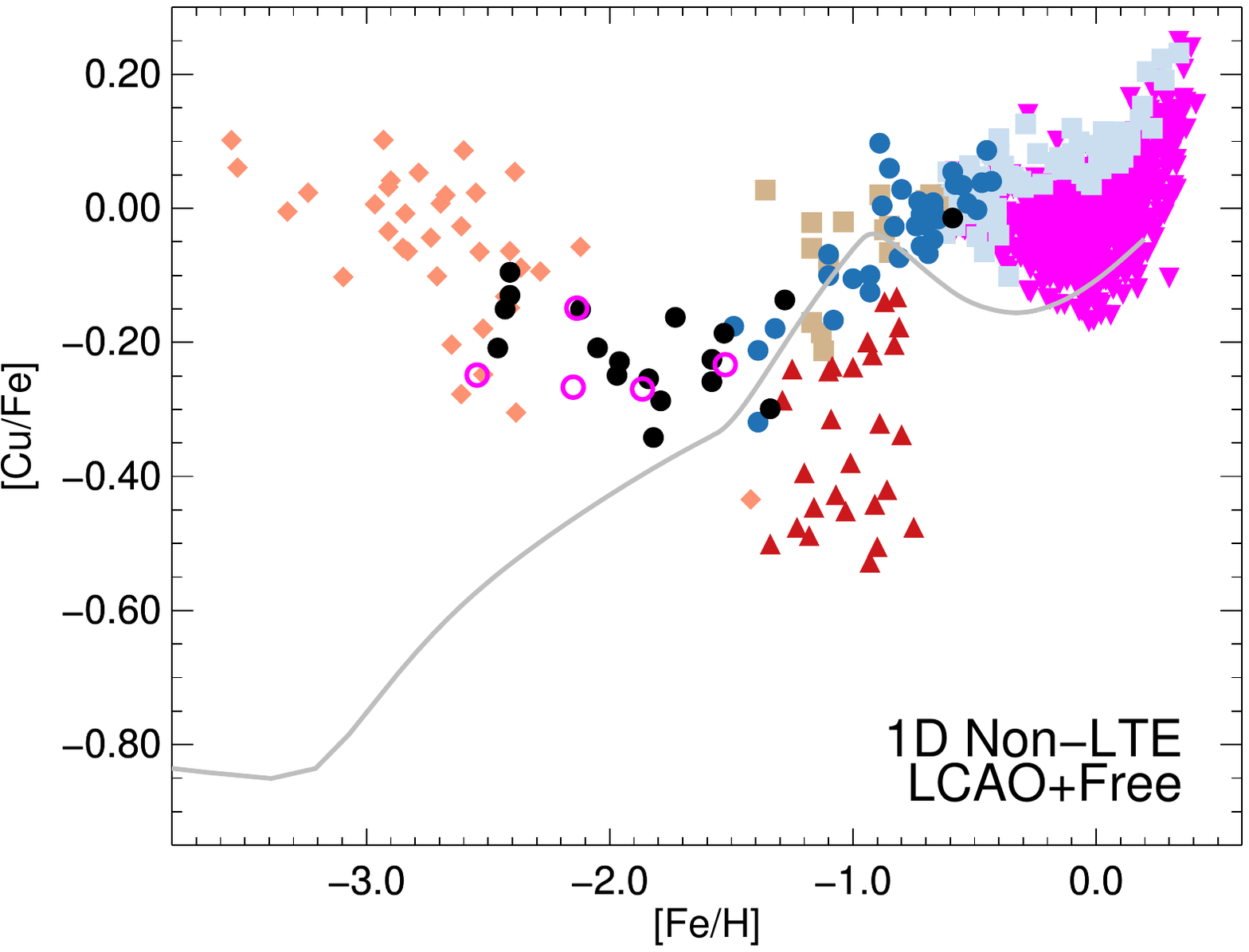}
        \caption{[Cu/Fe] versus [Fe/H]. 
        The top panel shows the LTE abundances from the literature, while the middle and bottom panels show non-LTE abundances using the hydrogen collision data from \cite{Belyaev_2021MNRAS.501.4968B} and our newly calculated LCAO+Free data, respectively. We again overplotted the same GCE model from \cite{2020ApJ...900..179K}. Literature data from: \citet{2017A&A...606A..94D} (dwarfs and subgiants in the thin disc  as magenta downwards triangles; thick disc and high-alpha metal-rich disc as light blue squares);
        \citet{Yan_2016A&A...585A.102Y} (dwarfs and subgiants in the thick disc as tan squares; high-alpha halo as dark blue circles; low-alpha halo as upwards red triangles);
        \citet{2018MNRAS.480..965K} (metal-poor dwarfs as magenta open circles); 
        \citet{2023ApJ...953...31S} (metal-poor dwarfs as light red diamonds);
        and \citet{2023MNRAS.518.3796C} (metal-poor giants as black filled circles).}
    \label{fig:copper_reanalysis}
\end{figure*}

The literature data described in \sect{litdata} is presented in \fig{fig:copper_reanalysis}. The first general observation is the significant difference between LTE and non-LTE trends. In LTE, the [Cu/Fe] trend follows the Galactic chemical evolution (GCE) model from \cite{2020ApJ...900..179K} more closely, exhibiting a nearly linear metallicity dependence in the metal-poor regime. A similar s-shaped [Cu/Fe] trend in LTE was reported by \cite{Bihain_2004A&A...423..777B} and \cite{Bisterzo_2004MmSAI..75..741B,Bisterzo_2005NuPhA.758..284B}, who found a linear increase in [Cu/Fe] for $-1.5 < \mathrm{[Fe/H]} < -1$, followed by a plateau at $\mathrm{[Fe/H]} < -1.8$. They argued that these results could be explained by a primary Cu production via core-collapse supernovae (CC SNe) at early times in our Galaxy, with secondary (and metallicity-dependent) production via the weak s-process in massive stars becoming increasingly important at later times \citep{Romano_2007MNRAS.378L..59R,Romano_2010A&A...522A..32R}. This nucleosynthesis scenario aligns with the recent analysis of \cite{2024A&A...682A.116N}, who shows $\mathrm{[Cu/Mg]}$ abundance trends in accreted and in-situ halo stars down to $\mathrm{[Fe/H]}\approx-1.8$, and thereby inferred that Cu is produced solely in massive stars during CC SNe, with no contribution from Type Ia supernovae.

In non-LTE, the trend deviates significantly from the GCE model towards low metallicities. At higher metallicities ($-0.5<\mathrm{[Fe/H]}<+0.4$), non-LTE effects and sensitivity to hydrogen collision data are the smallest, as expected. \citet{2017A&A...606A..94D} reported a wave-like trend in [Cu/Fe] at these metallicities, with a local maximum around [Fe/H] = $-$0.4, followed by a decline toward solar metallicities and a subsequent increase at super-solar metallicities. This pattern was later confirmed by \citet{2024A&A...684A..15N} for thin and thick disk giants. Our non-LTE results show a qualitatively similar trend, with the GCE model reproducing the shape reasonably well, albeit somewhat offset to lower [Cu/Fe] and [Fe/H].

At intermediate metallicities ($-$1.4 $<$ [Fe/H] $<$ $-$0.5), our non-LTE results confirm the previously observed dichotomy between high-$\alpha$ stars (tan squares and dark blue circles) with higher [Cu/Fe] and low-$\alpha$ stars (upward red triangles) with lower [Cu/Fe]. This separation, initially identified by \citet{2011A&A...530A..15N} and \citet{Yan_2016A&A...585A.102Y}, suggests that high-$\alpha$ halo stars share a similar chemical enrichment history with thick disk stars, while the low-$\alpha$ population likely formed in environments with slower star formation rates, where both Type Ia and CC SNe contributed to the chemical evolution, leading to increased scatter. \citet{2024A&A...682A.116N} confirmed that [Cu/Fe] increases linearly with [Fe/H] for the high-$\alpha$ population, consistent with Cu production through the weak s-process (via neutrons from the Ne-Mg reaction), while the low-$\alpha$ population exhibits greater scatter.

At the lowest metallicities ([Fe/H] $<$ $-$1.4), our new LCAO+Free hydrogen collision data results in an overall agreement between dwarfs and giants (magenta open circles and black circles in \fig{fig:copper_reanalysis}, respectively) that was not present in LTE or when using the B21 data. This agreement provides confidence in our non-LTE model. 

This consistency between dwarfs and giants has revealed a previously hidden feature in the [Cu/Fe] trend—a clear upturn toward lower metallicities, resulting in a local minimum around [Fe/H] = $-$1.7. This upturn and dip are not reproduced by the GCE model, nor reported previously in the literature. For example, \citet{Shi_2018ApJ...862...71S} reported a linear decrease in non-LTE [Cu/Fe] for $-$1 < [Fe/H] < $-$2.5, while \citet{Andrievsky_2018MNRAS.473.3377A} found a decrease from solar values down to [Cu/Fe] $\approx -0.35$ for $-$1.5 < [Fe/H] < $-$2.5, though their sample size was small.

\subsection{Implications on stellar yields and Galactic enrichment}

\begin{figure}[ht]
\centering
\includegraphics[width=\hsize]{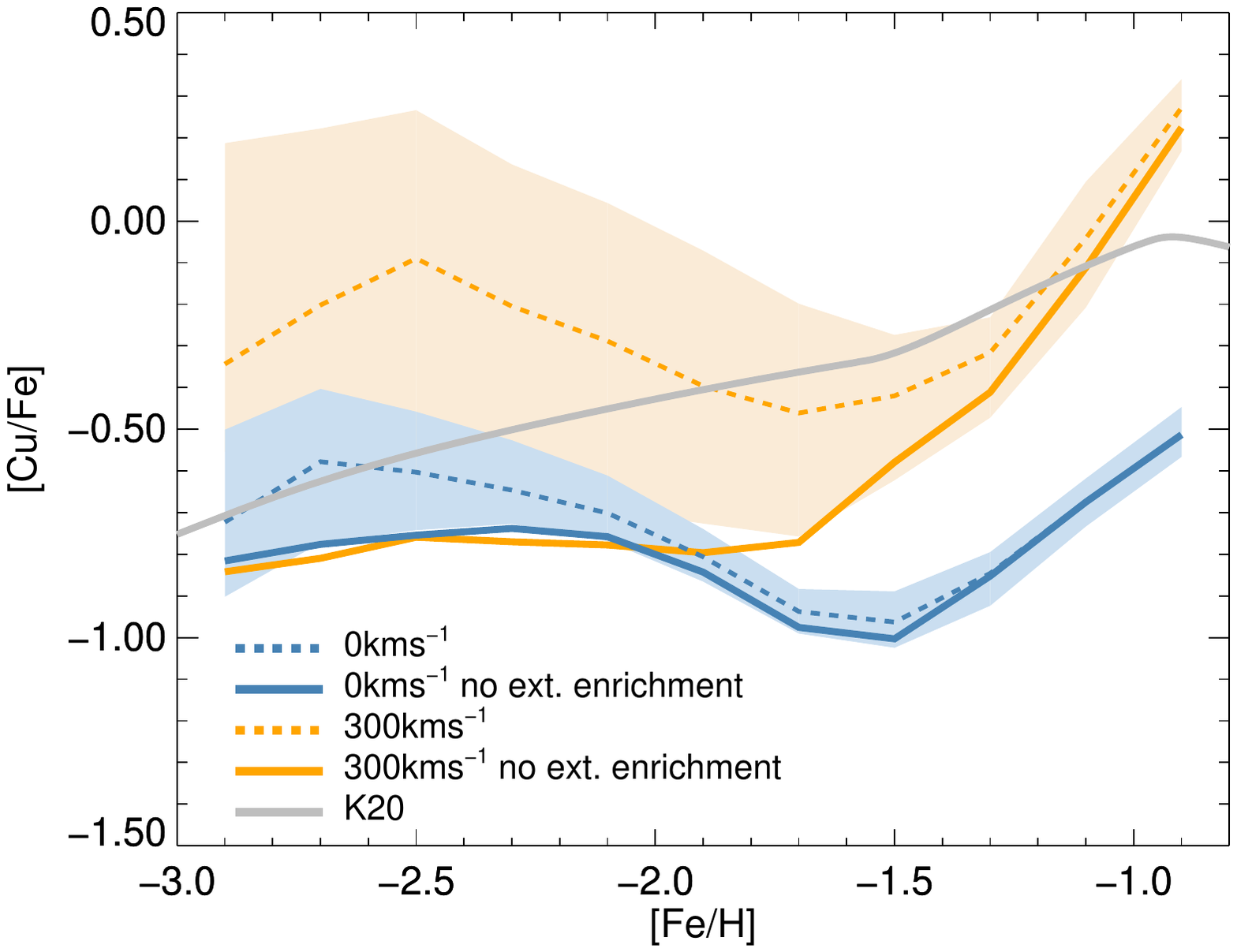}
\caption{\nefertiti{} models using Population~II yields from massive stars with two different rotation velocities \citep{Limongi_2018ApJS..237...13L}. The dotted lines and shaded areas represent the median and 25 and 75 percentiles of all stars in the model. The full lines represent the same models but only considering stars that formed in environments dominated by internal enrichment (no external enrichment case). The blue and orange colors are for the models using the non-rotating (v $=$ 0 km/s) and fast-rotating yields (v $=$ 300 km/s), respectively. The grey line is the GCE model from \cite{2020ApJ...900..179K}.
\label{fig:gce}}
\end{figure}

We can speculate on the origin of the upturn at low metallicity. One possibility is that it represents a PISN signature. In fact, in Figure 7 by \citet{Salvadori_2019MNRAS.487.4261S}, [Cu/Fe] versus [Fe/H] for an interstellar medium enriched by PISNe shows a probability peak around [Fe/H] = $-$1.7 for enrichment levels of 50 $\%$ and 90$\%$.

Additionally, we found that the yields of massive stars from \cite{Limongi_2018ApJS..237...13L} indicate a lower [Cu/Fe] at [Fe/H] $=$ $-$2 than at [Fe/H]= $-$3 and [Fe/H]= $-$1, creating a minimum around the same metallicities as our observed dip. However, the metallicity dependence of [Cu/Fe] yields varies with the initial rotation velocity of the progenitor massive star. For non-rotating stars, the minimum at [Fe/H] $\sim$ $-$2 is prominent, whereas increasing the rotation velocity diminishes this feature. 

In \fig{fig:gce}, we show the Cu abundance predictions from our cosmological GCE model \nefertiti{} adapted from the description in \cite{Koutsouridou_2023MNRAS.525..190K} and using the yields from \cite{Limongi_2018ApJS..237...13L} for Population~II stars. \nefertiti{} runs on N-body simulations of a Milky Way (MW) analogue and self-consistently accounts for the chemical enrichment of the gas in the MW progenitors and in the external ``MW environment'' or Intergalactic medium. 

In these models, we can see a dip in [Cu/Fe] at [Fe/H] $=$ $-$1.7 using both non-rotating and fast-rotating yields (see the shaded area and dotted lines in \fig{fig:gce}), although one would expect a flat trend for the fast-rotating case based on the behaviour of the yields described above.  The upturn at [Fe/H] $<$ $-$1.7 in our models arises because the stellar distribution at ${\rm [Fe/H]} \sim -3$ is dominated by stars formed in environments enriched by accretion of metal-rich intergalactic gas, contributing with a high [Cu/Fe] ratio.
Indeed, by considering only stars formed in halos dominated by internal enrichment, the \nefertiti{} trends follow the behaviour of the \cite{Limongi_2018ApJS..237...13L} yields, meaning a dip at [Fe/H] $\sim$ $-$1.7 with non-rotating yields, and a linear decrease of [Cu/Fe] with [Fe/H] followed by a plateau at [Fe/H] $<$ $-$1.7 with fast rotating yields (see solid lines in \fig{fig:gce}).

We also compared with the GCE from \cite{2020ApJ...900..179K}, which assumes only pristine gas accreting (i.e. no external enrichment), and found that it gives consistent results with the no external enrichment scenario from the \nefertiti{} model. In summary, the model accounting for external enrichment of the MW progenitors and using fast-rotating massive star yields appears to be reproducing our observed non-LTE abundances.

\section{Conclusion} \label{conclusion}

We have presented a new set of rate coefficients for Cu + H collisions based on an asymptotic LCAO approach, combined with rates from the free electron model for Cu for the first time. This way, we obtained a more complete dataset for hydrogen collisions and applied it to our Cu model atom to perform non-LTE calculations and derive Cu abundances.

By comparing our non-LTE results using our new hydrogen collision data with those obtained using previous data from the literature, we found that hydrogen collisions have a big impact on non-LTE corrections, particularly for hot, metal-poor stars. In metal-poor giants, in particular, we identified a large sensitivity to hydrogen collision data relative to the overall size of the non-LTE corrections.

These findings emphasize the need for accurate hydrogen collision rates for non-LTE models. Applying our new non-LTE Cu abundance corrections to LTE abundances in dwarfs and giants across a wide metallicity range, we observed a decrease in the line-to-line scatter of [Cu/Fe] in giants. Furthermore, our newly computed Cu + H collision data resulted in an improved agreement between Cu abundances in dwarfs and giants at a given metallicity bin. These astrophysical validations support the reliability of our newly calculated rates.

The non-LTE [Cu/Fe] versus [Fe/H] trend reveals an upturn in [Cu/Fe] for both dwarfs and giants at [Fe/H] $<$ $-$1.7, leading to a dip around [Fe/H] = $-$1.7 that is absent in LTE models and in non-LTE models using older hydrogen collision data. This feature may represent a potential PISN signature, where PISN-enriched material contributes to lower [Cu/Fe] values in the interstellar medium at this metallicity. If confirmed, this could provide an observational argument for the existence of PISN descendants. 

This dip in [Cu/Fe] may also indicate a combination of external enrichment through accretion of intergalactic gas and internal enrichment from fast-rotating massive stars, as suggested by the resemblance between the tested GCE models and the new non-LTE trends. This finding can help better constrain GCE models, and thus not only improve our understanding of stellar yields, but also uncover the metal-enrichment of the intergalactic gas and the build-up of the MW through this external enrichment channel.

Given the complexity of Cu nucleosynthesis, drawing firm conclusions from these new non-LTE trends remains challenging and additional work is needed. For example, the observed dip is primarily driven by a few metal-poor giants, so additional observations of giants at these metallicities can help confirm its significance. Alternatively, this pattern in [Cu/Fe] could result from uncertainties in the non-LTE model, such as electron collision data or $\log gf$ values. A full 3D non-LTE treatment is also necessary to rule out possible 3D effects. Finally, future work should explore additional GCE models with varying PISN contributions and yields to better understand the observed trends.

\section*{Data availability}

The Cu + H rate coefficients, line-by-line abundance corrections, and line-averaged [Cu/Fe] abundances are only available in electronic form at the CDS via anonymous ftp to cdsarc.u-strasbg.fr (130.79.128.5) or via http://cdsweb.u-strasbg.fr/cgi-bin/qcat?J/A+A/. The departure coefficients
are available at \url{https://doi.org/10.5281/zenodo.15062813}.

\begin{acknowledgements}
    AMA and PSB acknowledge support from the Swedish Research Council (VR 2020-03940 and VR 2020-03404).  AMA also acknowledges support from  the Crafoord Foundation via the
    Royal Swedish Academy of Sciences (CR 2024-0015).  
    MR and KL acknowledge funds from the European Research Council (ERC) under the European Union’s Horizon 2020 research and innovation programme (Grant agreement No. 852977) and funds from the Knut and Alice Wallenberg foundation.   
    This research was supported by computational resources provided
    by the Australian Government through the National Computational
    Infrastructure (NCI) under the National Computational Merit Allocation
    Scheme and the ANU Merit Allocation Scheme (project y89). The computations were also enabled by resources at the National Supercomputing Centre (NSC, Tetralith cluster) provided by the National Academic Infrastructure for Supercomputing in Sweden (NAISS), partially funded by the Swedish Research Council through grant agreement no. 2022-06725. We thank Elisabetta Caffau, Arthur Choplin, Adam Rains, and \'{A}sa Sk\'{u}lad\'{o}ttir for their assistance with data and the fruitful discussions. Finally, we thank the referee for their constructive comments, which helped improve the clarity of the manuscript.
\end{acknowledgements}

\bibliographystyle{aa_url} 
\bibliography{bibl.bib}

\begin{appendix}

\section{Calculation of parentage coefficients}\label{appendix}

\ion{Cu}{I} has states in the form of $3d^9 4s 4l$ $^2\mathrm{L}$, $^4\mathrm{L}$, where $l$ could be an $s$ or $p$ orbital. In the case of three mixed configurations like for the $3d^94s 4p~^2\mathrm{L}$, $^4\mathrm{L}$ states of \ion{Cu}{I}, the states of interest have the following coupling order:
\begin{IEEEeqnarray}{rCl}
(3d^n (S, L)_d \, \, 4s \, 4p (S, L)_{s+p}) (S, L)\,.\nonumber
\end{IEEEeqnarray}
However, we want to have the coupling order in the form of a core and a valence electron as such: 
\begin{IEEEeqnarray}{rCl}
(3d^n \, 4s (S, L)_c ) \, 4p)  (S, L)\,.\nonumber
\end{IEEEeqnarray}

To recouple the core, we need to recouple the angular part ($L_c$) and the spin part ($S_c$). The angular coupling is trivial since we have $l=0$ for $4s$, so: $L_c = L_d$.
For the spin recoupling, we have:
\begin{IEEEeqnarray}{l}
\label{eq:1}
    |((s_{s} s_{p})S_{s+p} \, S_d) SM\rangle =  
    \sum_{S_c} |(s_{s} (s_p s_d) S_c) SM\rangle \nonumber \\
     \times W(s_{s} s_p S \, S_d; S_{s+p} \, S_c) 
    \sqrt{(2S_{s+p}+1)(2S_c+1)}\,,
\end{IEEEeqnarray}
where $s_s = s_p = 1/2$ are the $s$ and $p$ electron spins, respectively, and $W$ is the Racah coefficient in the form of the Wigner 6-j symbol:
\begin{IEEEeqnarray}{c}
W(abcd;ef) = (-1)^{\text{a+b+c+d}} \begin{Bmatrix}
a & b & e\\
d & c & f
\end{Bmatrix}\,.\nonumber
\end{IEEEeqnarray}
By replacing $s_s$ and $s_p$ by 1/2, we get:
\begin{IEEEeqnarray}{l}
\label{eq:2}
|((\tfrac{1}{2} \tfrac{1}{2}) S_{s+p} S_d) SM\rangle = \sum_{S_c} |((\tfrac{1}{2}(\tfrac{1}{2} S_d) S_c) SM\rangle \nonumber \\
\times W(\tfrac{1}{2} \tfrac{1}{2} SS_d; S_{s+p} S_c) \sqrt{(2S_{s+p}+1)(2S_c+1)}\nonumber \\
= \sum_{S_c} |((\tfrac{1}{2}(\tfrac{1}{2} S_d) S_c) SM\rangle 
\begin{Bmatrix}
1/2 & 1/2 & S_{s+p}\\
S_d & S & S_c
\end{Bmatrix}
(-1)^{\tfrac{1}{2}+\tfrac{1}{2}+S+ S_d} \nonumber \\
\sqrt{(2S_{s+p}+1)(2S_c+1)}\,.
\end{IEEEeqnarray}

We have $S_{s+p} = s_s + s_p = 0 \, \, \mathrm{or} \, \, 1$, and $S= [ S_d - S_{s+p} ;S_d ;S_{d} +S_{s+p}]$.
Depending on the values of $S_{s+p}$ and $S$, we have only 2 cases where we have a mixing between two cores. Indeed, some combinations of $S_{s+p}$ and $S$ give us trivial cases where there is only one parent, so $G$'s are 1 and 0. Those combinations are $S_{s+p} = 0$ ; $S= S_d \pm 1$ and $S_{s+p} = 1$ ; $S= S_d \pm 1$ because for example if $S= S_d + 1$, $S_d = S -1$, so $S_c = S_d + s_s = S_d + 1/2 = S -1$. 

We can calculate the first case (referred to as ``case 1''), which corresponds to  $S_{s+p} = 0$ and $S= S_d$. Replacing $S_{s+p}$ and $S$ in \eqn{eq:2}, we have:
\begin{IEEEeqnarray}{l}
|((\tfrac{1}{2} \tfrac{1}{2}) S_{s+p} S_d) S M \rangle
= \sum_{S_c} |((\tfrac{1}{2}(\tfrac{1}{2} S_d) S_c) S_d M\rangle  \nonumber \\
\begin{Bmatrix}
1/2 & 1/2 & 0\\
S_d & S_d & S_c
\end{Bmatrix}
(-1)^{1+2S_d} \sqrt{(2S_c+1)} \, . \nonumber
\end{IEEEeqnarray}
Using
\begin{IEEEeqnarray}{c}
\begin{Bmatrix}
a & b & 0\\
d & e & f
\end{Bmatrix}= (-1)^{a+e+f} \frac{\delta_{ab} \delta_{de}}{\sqrt{(2a + 1)(2d + 1)}}\,, \nonumber
\end{IEEEeqnarray}
we get:
\begin{IEEEeqnarray}{l}
|((\tfrac{1}{2} \tfrac{1}{2}) S_{s+p} S_d) S M \rangle  
= \sum_{S_c} |((\tfrac{1}{2}(\tfrac{1}{2} S_d) S_c) S_d M\rangle \nonumber \\
\frac{(-1)^{\tfrac{1}{2} + S_d + S_c} (-1)^{1 + 2 S_d}}{\sqrt{2(2 S_d + 1)}} \sqrt{(2 S_c + 1)}\,. \nonumber
\end{IEEEeqnarray}
We can replace $S_c$ by $S_c = S_d \pm 1/2$ so that we obtain:
\begin{IEEEeqnarray}{l}
\label{eq:3}
|((\tfrac{1}{2} \tfrac{1}{2}) S_{s+p} S_d) S M \rangle 
= \sum_{S_d \pm 1/2} |((\tfrac{1}{2}(\tfrac{1}{2} S_d) S_c) S M\rangle \nonumber \\
(-1)^{4S_d + \tfrac{3}{2} \pm \tfrac{1}{2}} \frac{1}{\sqrt{2}} \sqrt{\frac{2S_d + 1 \pm 1}{2S_d + 1}}\,.
\end{IEEEeqnarray}

We can now do a similar development for the second case (i.e. ``case 2''), corresponding to $S_{s+p} = 1$ and $S= S_d$. Replacing $S_{s+p}$ and $S_c$ in \eqn{eq:2}, we have:
\begin{IEEEeqnarray}{l}
|((\tfrac{1}{2} \tfrac{1}{2}) S_{s+p} S_d) SM\rangle = 
\sum_{S_d \pm 1/2} |((\tfrac{1}{2}(\tfrac{1}{2} S_d) S_c) SM\rangle \nonumber \\
\begin{Bmatrix}
    1/2 & 1/2 & 1\\
S_d & S_d & S_d \pm1/2
\end{Bmatrix}
(-1)^{1+2 S_d} \sqrt{3(2S_c+1)}\,. \nonumber
\end{IEEEeqnarray}
After solving this, we obtain
\begin{IEEEeqnarray}{l}
\label{eq:4}
|((\tfrac{1}{2} \tfrac{1}{2}) S_{s+p} S_d) SM\rangle = 
 \sum_{S_d \pm 1/2} |((\tfrac{1}{2}(\tfrac{1}{2} S_d) S_c) SM\rangle \nonumber\\
\frac{1}{\sqrt{2}} \sqrt{\frac{2S_d + 1 \mp 1}{2S_d + 1}}\,.
\end{IEEEeqnarray}
So we obtain the same expression as \eqn{eq:3} in case 1, except now the $G$'s are always positive and the $\pm1$ is reversed with respect to $S_c = S_d \pm 1/2$.

We can now determine the $G$'s for \ion{Cu}{I} states. There are three cases we need to calculate:

\begin{enumerate}
  \item $3d^{9}(^2D)4s4p(^1P^o)~^2\mathrm{L^o}$ states:
  
These states belong to case 1 because $2S_{s+p} +1 = 1 \xrightarrow{} S_{s+p} = 0$. Since we have doublet states, $2S+1 = 2 \xrightarrow{} S = 1/2 = S_d$. 
So for the two ionic cores we have (using \eqn{eq:3}): 
\begin{IEEEeqnarray}{c}
S_c = S_d \pm 1/2 = 1 \, \, \mathrm{or} \, \, 0 \nonumber
\end{IEEEeqnarray}
\begin{IEEEeqnarray}{l}
3d^{9}4s~^3\mathrm{D} \xrightarrow{} S_c =1 \xrightarrow{} G = \frac{1}{\sqrt{2}} \sqrt{\frac{2S_d + 1 + 1}{2S_d + 1}} = 0.866 \nonumber
\end{IEEEeqnarray}
\begin{IEEEeqnarray}{l}
3d^{9}4s~^1\mathrm{D} \xrightarrow{} S_c =0  \xrightarrow{} G = - \frac{1}{\sqrt{2}} \sqrt{\frac{2S_d + 1 -1}{2S_d + 1}} = - 0.500 \nonumber
\end{IEEEeqnarray}

\item $3d^{9}(^2D)4s4p(^3P^o)~^2\mathrm{L^o}$ states:

These states belong to case 2 because $2S_{s+p} +1 = 3 \xrightarrow{} S_{s+p} = 1$. For the doublet states, we have $2S+1 = 2 \xrightarrow{} S = 1/2 = S_d$. 
So for the 2 ionic cores we have (using \eqn{eq:4}): 
\begin{IEEEeqnarray}{c}
S_c = S_d \pm 1/2 = 1 \, \, \mathrm{or} \, \, 0 \nonumber
\end{IEEEeqnarray}
\begin{IEEEeqnarray}{l}
3d^{9}4s~^3\mathrm{D} \xrightarrow{} S_c =1 \xrightarrow{} G = \frac{1}{\sqrt{2}} \sqrt{\frac{2S_d + 1 - 1}{2S_d + 1}} = 0.500  \nonumber
\end{IEEEeqnarray}
\begin{IEEEeqnarray}{l}
3d^{9}4s~^1\mathrm{D} \xrightarrow{} S_c =0  \xrightarrow{} G = \frac{1}{\sqrt{2}} \sqrt{\frac{2S_d + 1 +1}{2S_d + 1}} =  0.866 \nonumber
\end{IEEEeqnarray}

\item $3d^{9}(^2D)4s4p(^3P^o)~^4\mathrm{L^o}$ states:

We are still in case 2, but now for the quadruplet states. We have $2S+1 = 4 \xrightarrow{} S = 3/2 = S_d$. So for the 2 ionic cores we have: $$S_c = S_d \pm 1/2 = 2 \, \, \mathrm{or} \, \, 1$$ $$3d^{9}4s~^3\mathrm{D} \xrightarrow{} S_c =1$$ $$3d^{9}4s~^1\mathrm{D} \xrightarrow{} S_c =0$$ The latter case is impossible, so we have: 
\begin{IEEEeqnarray}{c}
G(3d^{9}4s~^3\mathrm{D}) = 1 \nonumber
\end{IEEEeqnarray}
\begin{IEEEeqnarray}{c}
G(3d^{9}4s~^1\mathrm{D}) = 0 \nonumber
\end{IEEEeqnarray}
\end{enumerate}

\end{appendix}

\end{document}